# Graph Searching, Parity Games and Imperfect Information[1]

Bernd Puchala and Roman Rabinovich

Mathematische Grundlagen der Informatik, RWTH Aachen University
{puchala,rabinovich}@logic.rwth-aachen.de

**Abstract.** We investigate the interrelation between graph searching games and games with imperfect information. As key consequence we obtain that parity games with bounded imperfect information can be solved in PTIME on graphs of bounded DAG-width which generalizes several results for parity games on graphs of bounded complexity. We use a new concept of graph searching where several cops try to catch multiple robbers instead of just a single robber. The main technical result is that the number of cops needed to catch $r$ robbers monotonously is at most $r$ times the DAG-width of the graph. We also explore aspects of this new concept as a refinement of directed path-width which accentuates its connection to the concept of imperfect information.

## 1 Introduction

The task of describing and modeling computing systems is intimately linked to interaction. Distributed computing devices, nonterminating reactive systems, knowledge bases and model checking all involve certain aspects of interaction. Many of these interactive situations take place under uncertainty: a controller does not necessarily have full information about the whole system state and the components of a distributed computing device do not have complete access to the implementation and actions of the other components. Furthermore, the model checking games for certain logics are games of imperfect information.

A model of interaction that has been studied extensively during the past decades is two-player games on graphs, especially infinite ones like parity games, see e.g. [6]. Parity games play a key role in modern approaches to verification and synthesis of state-based systems. They are the model-checking games for the modal $\mu$-calculus, a powerful specification formalism for verification problems that subsumes many temporal logics like LTL and CTL$^*$. Moreover, parity objectives can express all $\omega$-regular specifications and therefore capture fundamental properties of non-terminating reactive systems, cf. [13]. In these applications, the relevant problem is that of finding winning strategies for player 0.

For parity games with perfect information it is known that this problem is in NP ∩ co-NP [5] and it is an important open question whether it is in

[1] This work was partially supported by the ESF EUROCORES project LogiCCC, www.esf.org.

PTIME. However, it has been shown that the problem can be solved in PTIME on many classes of graphs of bounded complexity, e.g., on graphs of bounded Kelly-width [8], DAG-width [2] (and hence directed path-width), and entanglement [3]. On the other hand, finding winning strategies for player 0 in parity games with *imperfect information* is EXPTIME-complete in general [12] and it has been shown that this remains true for graphs of entanglement and directed path-width at most 2 (and hence DAG-width at most 3) [10].

Another natural restriction of the general setting is to bound the amount of uncertainty that player 0 has in the game. This is suitable, for example, for modeling reactive systems where the information of the controller, represented as player 0 in the game model, is acquired by sensors of a certain, though bounded, imprecision. Another possible source of bounded uncertainties is communication via noisy channels. For parity games with bounded imperfect information it has been shown that they can be solved in PTIME on graphs of bounded directed path-width [10]. For this, first, Reif's powerset construction [12] is applied to obtain a parity game with perfect information on a powerset graph which is only polynomially larger in the case of bounded imperfect information. Since parity games with perfect information can be solved in PTIME on graphs of bounded directed path-width, it remains to show that boundedness of this measure is preserved by the powerset construction. The approach from [10] uses the characterization of directed path-width via a cops and robber game on the given graph where several cops try to capture one robber on the graph *monotonously* (see Section 2.2). These graph searching games are also games of imperfect information themselves: the robber is invisible for the cops. As it turns out, this allows for a particularly easy translation of cops' strategies from the original to the powerset graph. In a sense, the imperfect information in the graph searching game captures the uncertainties of player 0 which are explicitly represented in the powerset graph.

This observation and the resulting fixed parameter tractability of parity games with bounded imperfect information give rise to a deeper analysis of the interrelation between graph searching games and the concept of imperfect information, especially in parity games. Here, we investigate the following aspect. To be able to translate strategies for the cop player from the original graph to the powerset graph, in the case of bounded imperfect information, it is not necessary to have a completely invisible robber. In fact, if $r$ is the maximal size of the subsets in the powerset graph, a robber which may be on at most $r$ possible vertices at each point, is sufficient. We formalize this idea by defining $\mathrm{dw}_r(G)$ as the number of cops needed to capture $r$ *visible* robbers monotonously on $G$ at the same time. Hence, the concept is both a refinement of directed path-width and a generalization of DAG-width, leading to a natural hierarchy $\mathrm{dw}(G) = \mathrm{dw}_1(G) \leq \mathrm{dw}_2(G) \leq \ldots \leq \mathrm{dw}_n(G) = \mathrm{dpw}(G)$ of complexity values, where dw denotes DAG-width, dpw denotes the directed path-width and $n$ is the number of vertices of $G$. We prove that this hierarchy does not collapse in general, thus obtaining a proper approximation of the directed path-width by means of bounded imperfect information.

The most important question that arises for this new concept of graph searching is, whether $\mathrm{dw}_r(G)$ for some given $r$ can be bounded by $\mathrm{dw}(G)$. Our main technical result is a positive answer to this question, more precisely we show that $\mathrm{dw}_r(G) \leq r \cdot \mathrm{dw}(G)$. To prove that $r$ robbers can be caught *simultaneously* and *monotonously* on a given graph the straightforward approach is to apply the given winning strategy $f$ against a single robber independently to the $r$ different robbers. For undirected graphs this simple idea in fact works, for directed graphs, however, the situation is more complicated, see Section 4.1.

A major problem here is that a cops' strategy against a single robber may place cops outside the robber's strongly connected component. This property of cops' strategies also gives rise to an example in [9] which shows that, in general, additional cops are necessary to convert a non-monotone winning strategy into a monotone one. Whether the number of additional cops is bounded is one of the most important open questions about DAG-width[2,9].

One possibility to solve this problem would be to translate the strategy $f$ into a certain normal form that would prescribe the cops to go inside the robber's component. Given the example from [9], this would also be a substantial step towards solving the monotonicity problem for DAG-width. However, in Section 4.1 we prove that for translating winning strategies into such a normal form unboundedly many additional cops are needed, so this approach does not work. Our solution is a more subtle way to apply the given (arbitrary) monotone winning strategy $f$ to $r$ robbers, see Section 4.2.

Finally, as key consequence of this analysis we obtain our second main result which states that parity games with bounded imperfect information can be solved in PTIME on graphs of bounded DAG-width. This generalizes the corresponding result for perfect information parity games from [2] to games with arbitrary, though fixed, amounts of uncertainty. Moreover, it generalizes the corresponding result for directed path-width from [10]. We also think that the techniques and insights established here can be applied to other problems on graphs which involve certain graph transformations like powerset or quotient constructions.

## 2 Preliminaries

For sets $X, Y \subseteq V$, the set $\mathrm{Reach}_{G-X}(Y)$ consists of vertices which are reachable from some $u \in Y$ via a path $P$ in $G$ such that $P \cap X = \emptyset$. For a finite sequence $\pi$ of some elements, $\mathrm{last}(\pi)$ denotes the last element of $\pi$. All graphs in this work are finite.

### 2.1 Parity games with imperfect information

In the applications mentioned in the introduction it is always sufficient to consider winning strategies only for player 0. Hence, our model of parity games with imperfect information has also imperfect information only for player 0, see also [11,10].

A parity game is a two-player game $G = (V, V_0, (E_a)_{a \in A}, \Omega)$ where $V$ is a finite set of positions, $V_0 \subseteq V$ is the set of positions of player 0 and $A$ is the finite set of actions. For each $a \in A$, $E_a \subseteq V \times V$ is the move-relation for action $a$ and $\Omega : V \to C \subsetneq \mathbb{N}$ is a coloring of $\mathcal{G}$ with colors from a finite set $C$. The game arena is the graph $(V, E)$ where $E = \bigcup_{a \in A} E_a$. A play is an infinite sequence $\pi = v_0 v_1 v_2 \ldots \in V^\omega$ of positions such that for each $i < \omega$ we have $(v_i, v_{i+1}) \in E$. A play $\pi$ is won by player 0 if the least color seen infinitely often in $\pi$ is even. A strategy for player 0 prescribes the next action for player 0 for any finite prefix of a play, which we also call history, where player 0 should move. So it is a function $g : \{\pi \in V^* \mid \text{last}(\pi) \in V_i\} \to A$. A play $\pi = v_0 v_1 v_2 \ldots$ is consistent with $g$ if for each $i < \omega$ with $v_i \in V_0$ we have $(v_i, v_{i+1}) \in E_{a_i}$ with $a_i = g(v_0 \ldots v_i)$. The strategy $g$ is called winning strategy for player 0 if each play that is consistent with $g$ is won by player 0.

A parity game with imperfect information $\mathcal{G} = (G, \sim)$ is given by a parity game $G = (V, V_0, (E_a)_{a \in A},)$ and an equivalence relation $\sim \subseteq V \times V$ which defines the vertices that are indistinguishable for player 0. We consider here only the case of parity games with observable colors, that means, if $u \sim v$ then $\Omega(u) = \Omega(v)$. A strategy for player 0 for $\mathcal{G}$ is a strategy $g$ for player 0 for $G$ which is based only on the information that player 0 has. Formally, for all finite histories $\pi = v_0 v_1 \ldots v_n$ and $\rho = w_0 w_1 \ldots w_n$ with $v_i \sim w_i$ for $i = 1, \ldots, n$ we require that $g(\pi) = g(\rho)$.

We say that a parity game $\mathcal{G} = (\mathcal{G}, \sim)$ has *imperfect information of size at most $r$* if $|[v]_\sim| = |\{u \in V \mid u \sim v\}| \leq r$, that means, the size of the largest equivalence class of positions is at most $r$. We say that a class of parity games has bounded imperfect information if there is some $r$ such that each game from that class has imperfect information of size at most $r$.

## 2.2 Graph searching games

A cops and robber game [2] $\mathcal{G}_k(G)$ is played on a directed graph $G = (V, E)$ by two players. The cops player controls $k$ cops where $k$ is a parameter of the game and the robber player controls a robber. Cops' positions are of form $(U, v)$ where $U \subseteq V$ is the set of at most $k$ vertices occupied by cops (if $|U| < k$, we say that the rest of the cops is outside of the graph) and $v \in V \setminus U$ is the vertex occupied by the robber. Robber's positions are of the form $(U, U', v)$ where $U$ and $v$ are as before and $U' \subseteq V$ is the set of at most $k$ vertices announced by the cops that will be occupied by them in the next position. From a position $(U, v)$, the cops can move to a robber's position $(U, U', v)$. From a position $(U, U', v)$, the robber can move to a cops' position $(U', v')$ where $v' \in \text{Reach}_{G-(U \cap U')}(v)$ and $v' \notin U'$. In the first move, the robber is placed on any vertex, i.e., the first move is $\bot \to (\emptyset, v)$ for any $v \in V$. Here $\bot$ is an additional dummy first position of any play.

The cops and multiple robbers games generalize the usual cops and robber games in that now, a number of cops tries to catch several robbers simultaneously instead of just a single robber. Let $G = (V, E)$ be a graph and $k, r \in \mathbb{N}$. The $k$ cops and $r$ robbers game $\mathcal{G}_k^r(G)$ is defined as follows. A position has the form $(U, R)$ or $(U, U', R)$ where $U, U', R \subseteq V$ with $|U|, |U'| \leq k$ and $|R| \leq r$. Here $U$ represents the vertices currently occupied by cops, $U'$ are vertices that the cops have announced to occupy in the next position and $R$ represents the vertices occupied by the robbers. From a cops' position $(U, R)$, the cops can move to any position $(U, U', R)$ as before. From a robbers' position $(U, U', R)$, the robbers can move to any position $(U', R')$ such that $R' \cap U' = \emptyset$ and each $r' \in R'$ is reachable from some $r \in R$ in $G - (U \cap U')$. In the first move, the robbers can go from $\bot$ to any position $(\emptyset, R)$ with $|R| \leq r$. Note that robbers can leave the graph. Furthermore there may be distinct $v_1, v_2 \in R'$ reachable only from the same vertex $v \in R$ in $G - (U \cap U')$. Informally, we say that robber $v_1$ *runs* and robber $v_2$ *jumps* if we assume that the robber on $v_1$ was on $v$ before the move and the robber on $v_2$ was on a vertex $w$ with $v_2 \notin \text{Reach}_{G-(U \cap U')}(w)$. Notice that this distinction is not made in the formalization.

A memory strategy for the cops player in a cops and (multiple) robber(s) game is is a memory structure $\mathcal{M} = (M, \text{init}, \text{upd})$ together with a function $f : M \times 2^V \times V \to 2^V$, resp. $f : M \times 2^V \times 2^V \to 2^V$. Hereby $M$ is a set of memory states, init $: V \to M$, resp. init $: 2^V \to M$ is the memory initialization function mapping the position after the first move of the robber(s) to a memory state, and upd $: M \times 2^V \times 2^V \times V \to M$, resp. upd $: M \times 2^V \times 2^V \times 2^V \to M$ is the memory update function, which maps a memory state and a cops' position to a new state. A memory strategy is positional if $|M| = 1$, in which case $M$ can be omitted. Winning strategies, plays, histories and consistency are defined for graph searching games in the usual way, analogously to the case of parity games, so we do not give formal definitions here. A play of a cops and (multiple) robber(s) game is *monotone* if it does not contain a position $(U, U', R)$ such that some $u \in U \setminus U'$ is reachable from some $r \in R$ in $G - (U \cap U')$. We also call a cops' strategy monotone, if every play consistent with it is monotone. A finite play is won by cops if it is monotone and there is no legal move for the robbers. Non-monotone plays are won by the robbers as well as infinite ones.

The minimal $k$ such that $k$ cops have a winning strategy for the monotone cops and $r$ robbers game on $G$ is denoted by $\text{dw}_r(G)$. The *DAG-width* of a graph $G$ is $\text{dw}_1(G)$. The notion of $\text{tw}_r(G)$ is defined in the same way as $\text{dw}_r(G)$, but the game is played on the graph $\overleftrightarrow{G} = (V, \overleftrightarrow{E})$ where $\overleftrightarrow{E} = \{(v, w) \mid (v, w) \in E \text{ or } (w, v) \in E)\}$, i.e., $\text{tw}_r(G) = \text{dw}_r(\overleftrightarrow{G})$. It is folklore that tree-width of a graph $G$, $\text{tw}(G)$ is equal to $\text{tw}_1(G) - 1$.

(Directed) path-width of a graph $G$ is the minimal number of cops that have a monotone [1] winning strategy against an *invisible* robber on $\overleftrightarrow{G}$ (on $G$). This is a game with imperfect information for the cop player where cops' strategies are functions $f$ that map sequences of cops' placements to a next placement: $f : (2^V)^* \to 2^V$.

When speaking about strongly connected components (SCCs) we shall refer to components in the graph $G - U$. For a vertex $v \in V$ we write $C(v)$ to denote the SCC $C$ with $v \in C$.

## 3 Parity games with bounded imperfect information

In this section, let $\mathcal{G} = (G, \sim)$ with $G = (V, V_0, (E_a)_{a \in A}, \Omega)$ be parity game with imperfect information (and observable colors) and let $\overline{G} = (\overline{V}, \overline{V}_0, (\overline{E}_a)_{a \in A}, \overline{\Omega})$ be the powerset graph of $\mathcal{G}$ according to Reif's construction ([12], see also [10]). Notice that vertices of $\overline{G}$ are sets of vertices of $G$, that means, $\overline{V} \subseteq 2^V$. We prove that, if $\mathcal{G}$ has imperfect information of size at most $r$, the DAG-width of $\overline{G}$ is bounded by $\mathrm{dw}_r(G) \cdot 2^{r-1}$. Together with our main technical result stating that $\mathrm{dw}_r(G) \leq r \cdot \mathrm{dw}(G)$ we can infer that parity games with bounded imperfect information can be solved in polynomial time on graphs of bounded DAG-width from the corresponding result for games with perfect information. We don't need the precise definition of $\overline{G}$ here but we use the following technical observation on the powerset construction which, while straightforwardly to prove, yields the key feature which allows to translate winning strategies for the cop player from the original game graph to the powerset graph.

**Lemma 1.** *For each finite history $\overline{\pi} = \overline{v}_0 \overline{v}_1 \ldots \overline{v}_n$ in $\overline{G}$ and all $v_n \in \overline{v}_n$, there is a finite history $\pi = v_0 v_1 \ldots v_n$ in $\mathcal{G}$ such that $v_i \in \overline{v}_i$ for all $i \in \{0, \ldots, n\}$.*

**Lemma 2.** *If $\mathrm{dw}_r(G) \leq k$ then $\mathrm{dw}(\overline{G}) \leq k \cdot 2^{r-1}$.*

*Proof.* Let $f$ be a winning strategy for the cops in $\mathcal{G}_k^r(G)$. We play a play of $\mathcal{G}_k^r(G)$ and a play of $\mathcal{G}_k(\overline{G})$ simultaneously and translate cops' moves from $\mathcal{G}_k^r(G)$ to $\mathcal{G}_k(\overline{G})$ and robber's moves vice versa. We maintain two invariants. The (Robbers) invariant is that if, in a position of $\mathcal{G}_k(\overline{G})$, the robber occupies a vertex $\overline{v} = \{v_1 \cdots, v_s\} \in \overline{V}$ with $s \leq r$ then, in the corresponding position in $\mathcal{G}_k^r(G)$ (after the same number of moves), the robbers occupy the set $\overline{v} \subseteq V$. The (Cops) invariant is that if the cops occupy a set $U$ in $\mathcal{G}_k^r(G)$ then, for every $u \in U$, the cops occupy every $\overline{u}$ in $\mathcal{G}_k(\overline{G})$ with $u \in \overline{u}$.

Assume the robber occupies a vertex $\overline{v} = \{v_1, \cdots, v_s\}$ with $s \leq r$ in $\mathcal{G}_k(\overline{G})$. We consider the robbers' move to $\{v_1, \cdots, v_s\}$ in $\mathcal{G}_k^r(G)$. To translate the cops' move, let $U' = f(U, \overline{v})$ be the cops' move in position $(U, \overline{v})$. We translate this move to $\mathcal{G}_k(\overline{G})$ as $\overline{U'}$ where $\overline{u} \in \overline{U'}$ if and only if $\overline{u} \cap U' \neq \emptyset$. For the robber's moves, consider a robber's position $(\overline{U}, \overline{U'}, \overline{v})$ in $\mathcal{G}_k(\overline{G})$ and a robber's move from $(\overline{U}, \overline{U'}, \overline{v})$ to $(\overline{U'}, \overline{w})$. Let $(U, U', \overline{v})$ and be the corresponding positions of $\mathcal{G}_k^r(G)$. We translate the robber's move to the move $(U, U', \overline{v}) \mapsto (U', \overline{w})$ in $\mathcal{G}_k^r(G)$. By (Cops), there is indeed a path from $\overline{v}$ to $\overline{w}$ in $G - (U \cap U')$. Using Lemma 1, it can be seen that the new strategy for $\mathcal{G}_k(\overline{G})$ is monotone. Moreover, it can be shown that the robber is finally caught.

To be more formal, consider any strategy $\overline{g}$ for the robber player for the monotone $k \cdot 2^{r-1}$ cops and (single) robber game on $\overline{G}$. We construct a play $\overline{\pi}_{fg}$ of this game that is consistent with $\overline{g}$ but not won by the robber player. As $\overline{g}$ is arbitrary, it follows that the cops have a winning strategy.

While constructing $\overline{\pi}_{fg}$ we simultaneously construct, for every finite prefix

$$\overline{\pi} = (\overline{U}_0, \overline{v}_0)(\overline{U}_0, \overline{U}_1, \overline{v}_0) \ldots (\overline{U}_{i-1}, \overline{U}_i, \overline{v}_{i-1})(\overline{U}_i, \overline{v}_i)$$

or
$$\overline{\pi} = (\overline{U}_0, \overline{v}_0)(\overline{U}_0, \overline{U}_1, \overline{v}_0) \ldots (\overline{U}_i, \overline{v}_i)(\overline{U}_i, \overline{U}_{i+1}, \overline{v}_i)$$

of $\overline{\pi}_{fg}$, a finite $f$-history

$$\zeta(\overline{\pi}) = (U_0, \overline{v}_0)(U_0, U_1, \overline{v}_0) \ldots (U_{i-1}, U_i, v_{i-1})(U_i, \overline{v}_i)$$

or

$$\zeta(\overline{\pi}) = (U_0, \overline{v}_0)(U_0, U_1, \overline{v}_0) \ldots (U_i, \overline{v}_i)(U_i, U_{i+1}, \overline{v}_i)$$

in the cops and $r$ robber game on $G$, such that for all $j \leq i$ we have

$$\overline{u} \in \overline{U}_j \text{ if and only if } \overline{u} \cap U_j \neq \emptyset.$$

Moreover, if $\overline{\pi}'$ is a prefix of $\overline{\pi}$ then $\zeta(\overline{\pi}')$ is a prefix of $\zeta(\overline{\pi})$.

First, with the history $\overline{\pi}$ which consists only of the initial move $(\emptyset, \overline{u})$ of the robber player, we associate $\zeta(\overline{\pi}) = (\emptyset, \overline{u})$. To translate the first cops' move, consider the set $U_0 = f(\zeta(\pi))$ of positions occupied by the cops in their first move according to $f$. We define $\overline{U}_0 = \overline{f}(\overline{\pi})$ by $\overline{u} \in \overline{U}_0$ if and only if $\overline{u} \cap U_0 \neq \emptyset$ and with $\overline{\pi}' = (\emptyset, \overline{u})(\emptyset, \overline{U}_0, \overline{u})$ we associate $\zeta(\overline{\pi}') = (\emptyset, \overline{u})(\emptyset, U_0, \overline{u})$.

For translating the robber's move in the induction step, consider any history $\overline{\pi} = (\overline{U}_0, \overline{v}_0)(\overline{U}_0, \overline{U}_1, \overline{v}_1)(\overline{U}_1, \overline{v}_2) \ldots (\overline{U}_{i+1}, \overline{v}_{i+1})$ and let, by induction hypothesis, $\zeta(\overline{\pi}(\leq i)) = (U_0, \overline{v}_0)(U_0, U_1, \overline{v}_0)(U_1, \overline{v}_1) \ldots (U_i, U_{i+1}, \overline{v}_i)$ be constructed. We define

$$\zeta(\overline{\pi}) = \zeta(\overline{\pi}(\leq i))(U_{i+1}, \overline{v}_{i+1})$$

and show that going from $\overline{v}_i$ to $\overline{v}_{i+1}$ is a legal robber's move in the game with $r$ robbers on $G$.

In the game on $\overline{G}$, the robber has just moved from $\overline{v}_i$ to $\overline{v}_{i+1}$, so $\overline{v}_{i+1} \notin \overline{U}_{i+1}$ and $\overline{v}_{i+1}$ is reachable from $\overline{v}_i$ in the graph $\overline{G}_{v_0} - (\overline{U}_i \cap \overline{U}_{i+1})$. Let $\overline{v}_i \xrightarrow{\overline{E}} \overline{v}^1 \xrightarrow{\overline{E}} \ldots \xrightarrow{\overline{E}} \overline{v}^t \xrightarrow{\overline{E}} \overline{v}_{i+1}$ be a path from $\overline{v}_i$ to $\overline{v}_{i+1}$ in $\overline{G} - (\overline{U}_i \cap \overline{U}_{i+1})$. Now let $v \in \overline{v}_{i+1}$. Then, by Lemma 1, there is some $u \in \overline{v}_i$ such that there is a path $u = u^0 \xrightarrow{E} u^1 \xrightarrow{E} \ldots \xrightarrow{E} u^t \xrightarrow{E} v$ in $G$ with $u^l \in \overline{v}^l$ for $l = 0, \ldots, t$. We have to show that $v \notin U_{i+1}$ and that $v$ is reachable from $u$ in $G - (U_i \cap U_{i+1})$. First, $\overline{v}_{i+1} \notin \overline{U}_{i+1}$ and therefore, by induction hypothesis for $\zeta(\overline{\pi}(\leq i))$, we have $\overline{v}_{i+1} \cap U_{i+1} = \emptyset$ which implies $v \notin U_{i+1}$. Now assume towards a contradiction, that $v$ is not reachable from $u$ in $G - (U_i \cap U_{i+1})$. Then there is some $l \in \{1, \ldots, t\}$ such that $u^l \in U_i \cap U_{i+1}$. But since $u^l \in \overline{v}^l$, by induction hypothesis for $\zeta(\overline{\pi}(\leq i))$, we have $\overline{v}^l \in \overline{U}_i \cap \overline{U}_{i+1}$ which contradicts the fact that $\overline{v}^1 \xrightarrow{\overline{E}} \ldots \xrightarrow{\overline{E}} \overline{v}^t$ is a path in $\overline{G} - (\overline{U}_i \cap \overline{U}_{i+1})$. Therefore, moving the robbers from $\overline{v}_i$ to $\overline{v}_{i+1}$ is a legal move for the robber player in the game with $r$ robbers on $G$, so $\zeta(\overline{\pi})$ is an $f$-history with the desired properties.

To translate the cops' answer, consider the set $U = f(\zeta(\overline{\pi}))$ of positions chosen by the cops to occupy in the next move according to $f$. We define $\overline{U} = \overline{f}(\overline{\pi})$ by

$$\overline{v} \in \overline{U} \text{ if and only if } \overline{v} \cap U \neq \emptyset,$$

that means, the cops occupy $\overline{v}$ if in the play on $G$ they occupy some vertex in $\overline{v}$. This yields the history $\overline{\pi}' = \overline{\pi}(\overline{U}_{i+1}, \overline{U}, \overline{v}_{i+1})$. With this history, we associate the history $\zeta(\overline{\pi}') = \zeta(\overline{\pi})(U_{i+1}, U, \overline{v}_{i+1})$ on $G$ which clearly has the desired properties.

We have to show that is won by the cops, i.e., that it is monotone and the robber is caught. To prove the monotonicity, assume, towards a contradiction, that the play $\overline{\pi}_{fg}$ is not monotone, i.e, there is a finite prefix $\overline{\pi} \prec \overline{\pi}_{fg}$ of $\overline{\pi}_{fg}$ such that $\text{last}(\overline{\pi}) = (\overline{U}_i, \overline{U}_{i+1}, \overline{v}_i)$ is a robber's position and such that there is some $\overline{u} \in \overline{U}_i \setminus \overline{U}_{i+1}$ which is reachable from $\overline{v}_i$ in $\overline{G} - \overline{U}_i \cap \overline{U}_{i+1}$. W.l.o.g. we can assume that there is a path $\overline{v}_i \xrightarrow{\overline{E}} \overline{v}^1 \xrightarrow{\overline{E}} \ldots \xrightarrow{\overline{E}} \overline{v}^t \xrightarrow{\overline{E}} \overline{u}$ from $\overline{v}_i$ to $\overline{u}$ in $\overline{G}$ with $\overline{v}^l \notin \overline{U}_i$ for $l = 1, \ldots, t$. Since $\overline{u} \in \overline{U}_i$ and $\overline{u} \notin \overline{U}_{i+1}$, by construction of $\zeta(\overline{\pi})$, we have $\text{last}(\zeta(\overline{\pi})) = (U_i, U_{i+1}, \overline{v}_i)$ and there is some $u \in \overline{u}$ with $u \in U_i$ and $u \notin U_{i+1}$. Moreover, by Lemma 1 there is some $v_i \in \overline{v}_i$ such that there is a path $v_i \xrightarrow{E} v^1 \xrightarrow{E} \ldots \xrightarrow{E} v^t \xrightarrow{E} u$ in $G$ with $v^l \in \overline{v}^l$ for all $l = 1, \ldots, t$. Hence $v^l \notin U_i$ for $l = 1, \ldots, t$ since if there is some $l \in \{1, \ldots, t\}$ such that $v^l \in U_i$ then by construction of $\zeta(\overline{\pi})$, we have $\overline{v}^l \in \overline{U}_i$ in contradiction to $\overline{v}^l \notin \overline{U}$ for $l = 1, \ldots, t$. So $u$ is reachable from $v_i$ in $G - U_i$. But since $v_i \in \overline{v}_i$ and $u \in U_i \setminus U_{i+1}$ and $(U_i, U_{i+1}, \overline{v}_i)$ occurs in a play which is consistent with $f$, this contradicts the fact that $f$ is strongly monotone.

Now assume that $\overline{\pi}_{fg}$ is won by the robber, i.e., $\overline{\pi}_{fg}$ is infinite. Then the play $\zeta(\overline{\pi}_{fg})$ which is obtained by combining all the finite histories $\zeta(\overline{\pi}_{fg}(\leq i))$ to

$$\text{last}\left(\left(\zeta(\overline{\pi}_{fg}(\leq 0)))\right)\right) \ldots \text{last}\left(\left(\zeta(\overline{\pi}_{fg}(\leq i)))\right)\right) \ldots$$

for $i < \omega$ is infinite as well. But since each history $\zeta(\overline{\pi}_{fg}(\leq i))$ is consistent with $f$, so is the play $\zeta(\overline{\pi}_{fg})$ which contradicts the fact that $f$ is a winning strategy for the cop player.

Finally, we count the number of cops used by the cops player in $\overline{\pi}_{fg}$. Consider any position $(\overline{U}_i, \overline{U}_{i+1}, \overline{v}_i)$ occurring in $\overline{\pi}_{fg}$. Since $\zeta(\overline{\pi}_{fg})$ is consistent with $f$, for the corresponding position $(U_i, U_{i+1}, \overline{v}_i)$ in $\zeta(\overline{\pi}_{fg})$ we have $|U_{i+1}| \leq k$ and by construction of $\overline{\pi}_{fg}$ it follows that $|\overline{U}_{i+1}| \leq k \cdot 2^{r-1}$. Therefore, the robber does not have a winning strategy against $k \cdot 2^{r-1}$ cops in the monotone cops and robber game on $\overline{G}$. By determinacy, $k \cdot 2^{r-1}$ cops have a winning strategy. □

**Theorem 3.** *[2] Parity games can be solved in polynomial time on graphs of bounded DAG-width.*

**Theorem 4.** *Parity games with bounded imperfect information can be solved in polynomial time on graphs of bounded DAG-width.*

*Proof.* Consider a class $\mathcal{K}$ of parity games $\mathcal{G} = (G, \sim)$ with bounded partial information and bounded DAG-width. Let $r$ be the maximal size of $\sim$-equivalence classes in games from $\mathcal{K}$ and let $k$ denote the maximal DAG-width of the corresponding game graphs. By Theorem 10, for any game $\mathcal{G}$ from $\mathcal{K}$ we have $\text{dw}_r(G) \leq k \cdot r$ and hence, by Lemma 2, $\text{dw}(\overline{G}) \leq k \cdot r \cdot 2^{r-1}$. Therefore, by applying the powerset construction to the games from $\mathcal{K}$ we obtain a class $\overline{\mathcal{K}}$ of parity games with perfect information which have bounded DAG-width. By

Theorem 3, the games from $\overline{\mathcal{K}}$ can be solved in polynomial time. Moreover, as $r$ is fixed, the size of the powerset games $\overline{G}$ from $\overline{\mathcal{K}}$ is polynomial in the size of of the original games from $\mathcal{K}$, so the games from $\mathcal{K}$ can be solved in polynomial time as well. □

## 4 From one robber to $r$ robbers

We say that a robbers' strategy $g$ is *isolating* if in any cops' position $(U, R)$ of a play that is consistent with $g$, for all $v, w \in R$, we have $v \notin \text{Reach}_{G-U}(w)$. In particular, two robbers never stay in the same SCC. It is easy to see that this is not a substantial restriction: the robber from the smaller vertex $v$ is redundant. He can still go to his current position in the next move by first jumping to the robber from a longer history on $v'$ and then running from $v'$ to $v$.

**Lemma 5.** *If $r$ robbers have a winning strategy against $k$ cops then $r$ robbers have an isolating winning strategy against $k$ cops.*

*Proof.* Given a set of vertices $U$, we say that $R$ and $\hat{R}$ are equivalent, $R \equiv_U \hat{R}$, if for all $r \in R$ there is some $\hat{r} \in \hat{R}$ and vice versa, for all $\hat{r} \in \hat{R}$ there is some $r \in R$ such that $r$ and $\hat{r}$ are in the same component of $G - U$.

Let $f$ be a winning strategy for $r$ robbers in the monotone multiple robbers game on a graph $G$ against $k$ cops. We construct a strategy $\hat{f}$ for $r$ robbers against $k$ cops by induction on the play length and show simultaneously the following. For each play $\pi$ which is consistent with $f$ there is a play $\hat{\pi}$ which is consistent with $\hat{f}$ (and conversely, for each $\hat{\pi}$ there is some $\pi$), the reachability regions of all robbers in both plays are the same. In other words, if $(U, U', R) \to (U', R')$ is the $i$-th robbers' move in $\pi$ and if $(U, U', \hat{R}) \to (U', \hat{R}')$ is the $i$-th robbers' move in $\hat{\pi}$ then $\text{Reach}_{G-U'}(R') = \text{Reach}_{G-U'}(\hat{R}')$. This is achieved as follows. Consider the topological order on vertices of $G - U'$. If $f$ prescribes to move from $(U, U', R)$ to $(U', R')$ then $\hat{f}$ prescribes to move from $(U, U', \hat{R})$ to $(U', \hat{R}')$ where $\hat{R}'$ is a set of topologically minimal vertices of $R'$ such that $\hat{R}'$ contains only one vertex from any equivalence class of $\equiv_{U'}$. We have to show that (1) such a move is possible, i.e., $R' \subseteq \text{Reach}_{G-U'}(\hat{R})$ and (2) that the invariant $\text{Reach}_{G-U'}(R') = \text{Reach}_{G-U'}(\hat{R}')$ holds. Condition (1) follows directly from the induction hypothesis, that is from $\text{Reach}_{G-U'}(R) = \text{Reach}_{G-U'}(\hat{R})$ because $R' \subseteq \text{Reach}_{G-U'}(R)$ and condition (2) is clear by construction of $\hat{f}$. □

### 4.1 Tree-width and componentwise hunting

Our main technical result states that to catch several robbers monotonously on a given graph, the number of needed cops is only increased by a factor which is equal to the number of robbers. As a start, we first consider the same result for the game characterizing tree-width.

**Lemma 6.** *For all $G$ and $k, r > 0$, if $\text{tw}(G) \leq k$ then $\text{tw}_r(G) \leq r \cdot (k+1)$.*

*Proof.* Let $f$ be a monotone winning strategy for $k$ cops in the game on $\overleftrightarrow{G}$ against one robber. As $f$ is monotone, we can assume that cops are not placed on vertices that are already unavailable for the robber, i.e., for a move $(U, v) \to (U, U', v)$ we always have $U' \setminus U \subseteq \text{Reach}_{G-(U \cap U')}(v)$. We construct a monotone strategy $\otimes_r f$ for $k \cdot r$ cops in the game on $\overleftrightarrow{G}$ with $r$ robbers that is winning against each isolating robbers' strategy.

Intuitively, the cop player uses $r$ teams of cops with $k$ cops in each team. Every team plays independently of each other chasing its own robber according to $f$. We maintain the invariant that in each cops' position $(U, R)$ that is consistent with $\otimes_r f$, there is a partition $(U_1, \cdots, U_r)$ of $U$ and an enumeration of $v_1, \cdots, v_r$ of $R$ such that for each $v_i$, $(U \setminus U_i) \cap \text{Reach}_{G-U_i}(v_i) = \emptyset$, i.e., cops on $U_i$ block $v_i$ from other cops and that $(U_i, v_i)$ is consistent with $f$ in the game with one robber. The next move of the cops is $\otimes_r f(U, R) = \bigcup_{i=1}^{r} f(U_i, v_i)$. By a simple induction on the length of a play it is easy to see that the invariant holds, which implies that the cops monotonously catch all $r$ robbers. □

The reason why the proof is so simple is that in an *undirected* graph the set of vertices which is reachable from a given position is precisely the connected component which contains this positions, so the strategy $f$ does not need to place cops on vertices outside the robber component. For directed graphs, this is not true and the simple translation of strategies is not possible without certain refinement any more. Consider the following possible situation. The cops play simultaneously against all robbers according to a winning strategy $f$ for the game against one robber as before. Alternatively, they choose one of them (occupying a vertex $v_1$) to play against him further while the cops of other teams wait for this robber to be caught. (This will be our approach in the proof of Theorem 10.) The robbers stay in two distinct SCCs on $v_1$ and $v_2$. The problem is that $v_2$, can prevent playing against $v_1$. If $f$ says to place a cop on a vertex $v$ that is reachable from $v_2$, it may become impossible to reuse the cop from $v$ later playing against $v_1$, although $f$ prescribes to do so: $v_2$ would induce non-monotonicity on $v$. Our solution is to *omit* to place the cop on $v$ and to play against $v_1$ further according to $f$. The cops from the team of $v_2$ have the duty to guard every vertices that is not guarded by the (absent) cop on $v$. If robber on $v_2$ leaves his vertex and jumps (say, to $v_1$), the cops from his team play according to $f$ from the position they stopped until they occupy $v$. Thus the omitted move to $v$ is performed later.

Notice that there is another, more straightforward, approach to solve this problem: to change $f$ such that it does not prescribe to place cops outside of the robber's component It would suffice to prove that there is a function $F: \mathbb{N} \to \mathbb{N}$ such that every cops' winning strategy $f$ for $k$ cops against one robber can be transformed into a winning strategy $f'$ for $F(k)$ cops against one robber that never prescribes to place cops outside of the robber's SCC. In other words, strategy $f'$ should fulfill the following property: in a position $(U, v)$, if $C$ is the SCC of $G - U$ with $v \in C$ then $f'(U, v) \subseteq C$. However, such a function $F$ does not exist.

**Theorem 7.** *There are graphs $G_n$, $n \in \mathbb{N}$, such that $\mathrm{dw}(G_n) \leq 4$ for all $n \in \mathbb{N}$, but any winning strategy of the cop player which is restricted to place cops only inside the robber's SCC, uses at least $n + 1$ cops.*

*Proof.* Consider the following class of directed graphs (see Figure 1). Every graph $G_n = (V_n, E_n)$ for $0 < n < \omega$ is an undirected full tree $\mathcal{T}_n = (T_n, B_n)$ of degree and depth $n+1$ together with another tree $\mathcal{T}'_n = (T'_n, B'_n)$ of the same shape with edges directed to the root. That means, $T_n = \{1, \ldots, n\}^{\leq n+1}$ and $B_n$ contains edges $(v, vj)$ and $(vj, v)$ for any $v \in \{1, \ldots, n\}^{\leq n}$ and any $j \in \{1, \ldots, n\}$. Further, $T'_n = \{1', \ldots, n'\}^{\leq n+1}$ and $B'_n$ contains edges $(vj, v)$ for any $v \in \{1', \ldots, n'\}^{\leq n}$ and any $j \in \{1, \ldots, n\}$. Additionally, from any vertex $v_1 \ldots v_m \in T_n$ of the first tree there is an edge to the corresponding vertex $v'_1 \ldots v'_m \in T'_n$ of the second tree and from any vertex $v'_1 \ldots v'_m v'_{m+1} \in T'_n$ of the second tree, there is an edge to the corresponding parent vertex $v_1 \ldots v_m \in T_n$ of the first tree.

It is easy to see that four cops capture a robber on every such graph by searching both trees $(T_n, E_n)$ and $(T'_n, E'_n)$ in a top-down manner in parallel. We show that, on $G_n$, the robber can defeat $n$ cops who do not place themselves outside his SCC. Assume that the cops occupy some set $U \subseteq V_n$ and the robber is on some vertex $v = v_1 \ldots v_m \in T_n$ such that the following invariant holds.

(1) Any strict ancestor $w \prec v$ of $v$, $w \in T_n$ is occupied by a cop, and
(2) any ancestor $w' \preceq v'$ of the corresponding vertex $v' = v'_1 \ldots v'_m$ is cop-free.

Note that due to condition (1), none of the vertices $w' \preceq v'$ lies in the SCC $C(v)$ of $v$. (In $G_n - U$, the only successor of a vertex $w' = v'_1 \ldots v'_r v'_{r+1} \preceq v'$ is $v'_1 \ldots v'_r$, so there is no path from $w'$ to $v$ in $G_n - U$.)

Assume that the cops move from $U$ to some $S$. As they do not place themselves outside of $C(v)$, they cannot occupy any ancestor $w' \preceq v'$, i.e., for $\mathrm{Pre}(v') = \{w' \in T'_n \mid w' \preceq v'\}$, we have $\mathrm{Pre}(v') \cap S = \emptyset$. Consider two cases. If there is some $w \prec v$ such that $w \notin S$ then the robber runs to the minimal (w.r.t. $\preceq$) such $w = v_1 \ldots v_r$ via the cop free path $v \to v' = v'_1 \ldots v'_m \to v'_1 \ldots v'_{m-1} \to \ldots \to v'_1 \ldots v'_r v'_{r+1} \to v_1 \ldots v_r$. Due to the choice of $w$ and the fact that $\mathrm{Pre}(v') \cap S = \emptyset$, the robber is then on some vertex $v \in T_n$ such that conditions (1) and (2) hold. In the second case, if there is no such $w$, then due to condition (1) and the fact that there are at most $n$ cops, $|v| \leq n$. If $v \notin S$ then the robber remains on $v$ and, obviously, conditions (1) and (2) hold. If $v \in S$ then due to the fact that there are at most $n$ cops, there is at least one $j \in \{1, \ldots, n\}$ such that the whole subtree rooted in $vj$ (including all the corresponding vertices from $T'_n$) is cop-free and the robber moves to $vj$. So the robber is again on some vertex $v \in T_n$ such that conditions (1) and (2) hold. Hence, the robber is never captured. □

### 4.2 Generalization to the directed case

In this section we prove our main technical result. For this, we need some additional notions and lemmata. First, we introdue prudent strategies: a strategy for the robber player is called *prudent* if, according to this strategy, a robber runs from his current vertex to another one, only if staying at the current vertex would make the target vertex unavailable for the robbers after the cops have landed. Formally, the moves $(U, U', R) \to (U', R')$ of the robber player are restricted by the condition that any $r' \in R' \setminus R$ is not reachable in $G - U'$ from $R$.

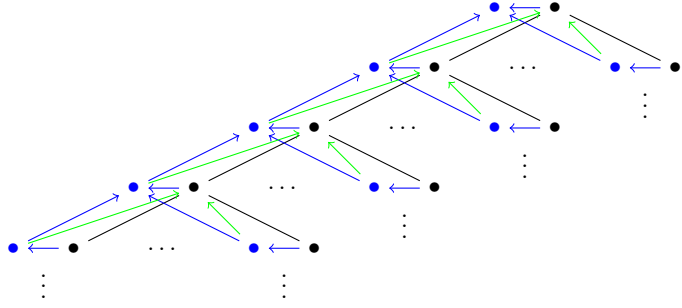

**Fig. 1.** $\mathrm{dw}(G_n) = 4$, but the robber wins against $n$ cops if they move only into his component.

The proof of the following lemma is very similar to the proof of Lemma 5 and we omit it. The difference is that the invariant $\mathrm{Reach}_{G-U'}(R') = \mathrm{Reach}_{G-U'}(\hat{R}')$ is replaced by the invariant $\mathrm{Reach}_{G-U'}(R') \subseteq \mathrm{Reach}_{G-U'}(\hat{R}')$.

**Lemma 8.** *If $r$ robbers have a winning strategy against $k$ cops then $r$ robbers have an isolating prudent winning strategy against $k$ cops.*

In the following lemma we show that any positional cops' winning strategy for game with one robber can be modified without using additional cops to obtain a new positional strategy that does not place a cop on a vertex that is already unavailable for the robber and always prescribes to place new cops.

**Lemma 9.** *On a graph $G$, if $f$ is a positional monotone winning strategy for $k$ cops against one robber then there is a positional monotone winning strategy $\tilde{f}$ for $k$ cops against one robber, such that for any finite history $\pi'$ consistent with $\tilde{f}$, if $\mathrm{last}(\pi') = (U, v)$, we have $\tilde{f}(\pi') \setminus U \neq \emptyset$ and any $u \in \tilde{f}(\pi') \setminus U$ is reachable from $v$ in $G - U$.*

*Proof.* We first construct a strategy $\hat{f}$ that never places a cop on a vertex that is already unreachable for the robber and then construct from $\hat{f}$ a strategy $\tilde{f}$ that, in addition, never prescribes the cops to stay idle or only to leave the graph.

The new strategy $\hat{f}$ is constructed from $f$ by induction on the length of play prefixes. Simultaneously we show two invariants. The first is that for all plays $\pi$ consistent with $f$ there is a play $\hat{\pi}$ consistent with $\hat{f}$ and vice versa (for any $\hat{\pi}$ there is some $\pi$) such that for all lengths $i$ of play prefixes, $\pi(i) = (U, U', v)$ if and only if $\hat{\pi}(i) = (\hat{U}, \hat{U}', v)$ and $\pi(i) = (U, v)$ if and only if $\hat{\pi}(i) = (\hat{U}, v)$ such that $\hat{U} \subseteq U$, $\hat{U}' \subseteq U'$ and for all $\hat{u} \in \hat{U}$, $\hat{u} \operatorname{Reach}_{G-(\hat{U}-u)}(v)$ and $\hat{u}' \in \hat{U}'$, $\hat{u}' \operatorname{Reach}_{G-(\hat{U} \cap \hat{U}'-u')}(v)$. The second invariant is that $\operatorname{Reach}_{G-(U \cap U')}(v) = \operatorname{Reach}_{G-(\hat{U} \cap \hat{U}')}(v)$. Thus $\pi$ is won by the cops if and only if $\hat{\pi}$ is won by the cops and hence $\hat{f}$ is winning.

The strategy $\hat{f}$ is defined as follows. Assume a cops' position $(\hat{U}, v)$. Then according to the first invariant, there is a position $(U, v)$ with $\hat{U} \subseteq U$ that occurs in a play consistent with $f$. Take an arbitrary such $(U, v)$ and let $f(U, v) = U'$. Then $\hat{f}(\hat{U}, v) = \hat{U}'$ where $\hat{U}' = \{\hat{u}' \mid \hat{u}' \operatorname{Reach}_{G-(\hat{U}-\hat{u}')}\}$. It is clear that the invariants hold and that $\hat{f}$ is positional.

Now, from $\hat{f}$, we construct strategy $\tilde{f}$ that, in addition to the properties of $\hat{f}$, in each move places at least one cop on the graph. Assume a position $(\hat{U}, v) = \operatorname{last}(\hat{\pi})$ with some finite play prefix $\hat{\pi}$ where $\hat{f}$ does not prescribe to place any cops. Thus $\hat{f}(\hat{U}, v) = \hat{U}_0$ where $\hat{U}_0 \subseteq \hat{U}$. Consider the prolongation of the play where the robber does not move, i.e., $\hat{\pi} \cdot (\hat{U}_0, v) \cdot (\hat{U}_1, v) \cdot \ldots$ where $\hat{U}_i \subseteq \hat{U}_{i+1}$ for all $i \geq 0$. As $\hat{f}$ is winning, there is a natural number $i$ such that $\hat{f}(\hat{\pi} \cdot (\hat{U}_0, v) \cdot \ldots \cdot (\hat{U}_i, v)) = \hat{U}'$ where $\hat{U}' \not\subseteq \hat{U}$, i.e., a cop is finally placed outside of $\hat{U}$. (Otherwise the robber will always stay on $v$ and no cop will occupy $v$.) Then define $\tilde{f}(\hat{U}, v) = \hat{U}'$. It is obvious that any play $\tilde{\pi}$ consistent with $\tilde{f}$ corresponds to a play $\hat{\pi}$ consistent with $\hat{f}$ such that one can obtain $\tilde{\pi}$ by cutting off some positions from $\hat{\pi}$. Therefore $\tilde{f}$ is winning and never places cops on vertices unreachable for the robber. Further, there are no idle moves according to $\tilde{f}$ by construction. Finally, $\tilde{f}$ is positional. □

With these normal forms for cops' and robbers' strategies at hand we can prove the following result.

**Theorem 10.** *For $k, r > 0$, if $\operatorname{dw}(G) \leq k$ then $\operatorname{dw}_r(G) \leq k \cdot r$.*

To prove this theorem, let $f$ be a positional monotone winning strategy for the cop player for the $k$ cops and (one) robber game on a directed graph $G$. According to Lemma 9 we can assume w.l.o.g. that for any finite history $\pi'$ consistent with $\tilde{f}$ such that $\operatorname{last}(\pi') = (U, v)$ we have $\tilde{f}(\pi') \setminus U \neq \emptyset$ and any $u \in \tilde{f}(\pi') \setminus U$ is reachable from $v$ in $G - U$. Moreover, due to Lemma 8 it suffices to construct a strategy $\otimes_r f$ for the cop player for the $r \cdot k$ cops and $r$ multiple robbers game on $G$ which is winning against all isolating prudent strategies for the robber player. First, we only sketch a description of a memory strategy $\otimes_r f : \mathcal{M} \times (2^V \times 2^V) \to 2^V$ and the corresponding memory structure.

The cops play in $r$ teams à $k$ cops. Consider a position $(U, R)$ in a play with $r$ robbers. With every vertex $v \in R$ that is occupied by a robber, we associate a team of cops $U_i \subseteq V$ with $|U_i| \leq k$. Note that some $U_i$ may coincide and we identify them. For each $U_i$ we associate a history $\rho_i$ of the game against one robber that is consistent with $f$ such that $(U_i, v)$ is the last position of $\rho$. We formulate this as an invariant in the game with $r$ robbers:

**(Cons)** Any history $\rho_i$ is consistent with $f$.

Let $\prec$ be the (irreflexive) prefix relation on finite histories of the game with one robber seen as words of consecutive positions. We keep at most $r$ histories $\rho_i$ in memory and write $\rho = \rho_1, \cdots, \rho_s$ for some $s \leq r$. This sequence of histories is the main part of the memory. The following invariant says that, up to the last robber's moves, all $\rho_i$ are linearly ordered by $\prec$.

**(Lin)** $\rho_1 \prec \rho_2 \prec \ldots \prec \rho_s$.

The sequence $\rho$ is constructed and maintained in the memory in the following way. At the beginning of a play, we set $\rho = \rho_1 = \bot$. Now consider the maximal play prefix $\bot (U^1, R^1) \cdots (U^m, R^m)(U^m, (U^m)', R^m)$ in the game with $r$ robbers where all $R^i$ are singletons. While playing this part of the play, all teams make the same moves according to $f$. We save the sequence as $\rho = \rho_1 = \bot (U^1, v^1) \cdots (U^m, v^m)(U^m, (U^m)', v^m)$ where $\{v^i\} = R^i$ (see Figure 2). When the robbers go into different SCCs, the cops choose one of them, say on a vertex $b_1$. Let the set of vertices occupied by other robbers be $R_1$. We associate $\rho_2 = \rho_1(U^m, (U^m)', b_1)$ and store $\rho = \rho_1, \rho_2$. Note that $\rho_1$ ends with a robber's position. Assume for a moment that only the robber in $C(b_1)$ moves. Then only this robber is pursued according to $f$, but cops are not placed on vertices $v$ for any $v \in \text{Reach}_{G-U^m} R_1$. These moves are appended to $\rho_2$, however without respecting the omitted placements. Formally, let $W_2$ be the last cops placement in $\rho_2$ and let $b_2$ be the last robbers' vertex in $\rho_2$. Then, in a position $(U, R)$ of the game with $r$ robbers, we have $\otimes_r f(U, R) = f(W_2, b_2) \setminus \text{Reach}_{G-W_1}(b_2)$. (Note that $\otimes_r f$ depends also on the memory state, but we will not write it explicitly.) In $\rho_2$, not the actual moves $f(W_2, b_2) \setminus \text{Reach}_{G-W_1}(b_2)$ are stored, but the intended one, i.e., $f(W_2, b_2)$. If later new robbers come and occupy different SCCs of $C(b_2)$ we again choose one of them (on $b_3 \in V$), create $\rho_3$ and set $\rho_3$, $W_3$ and $b_3$ analogously to $\rho_2$, $W_2$ and $b_2$, and $\rho = \rho_1, \rho_2, \rho_3$. The cops play according to $\otimes_r f(U, R) = f(W_3, b_3) \setminus \text{Reach}_{G-W_2}(b_2)$.

Note that histories in $\rho$ are subject to change, so at different points of time, $\rho$ and $\rho_i$ are different objects. Note further that cops from other teams smaller than 3 (in general, $s$) cannot be taken from their vertices, as, according to $f$, omitted placements must be performed first, so taking the cops may infer non-monotonicity. Note also that there may be more than one robber in $R_i$ associated to a play $\rho_i$ if $i < s$ and at most one robber is associated with the longest history.

A complete element of the memory structure has the form

$$\zeta = (\rho_1, R_1, O_1), \ldots, (\rho_{s-1}, R_{s-1}, O_{s-1}), \rho_s.$$

Hereby the $\rho_i$ and $R_i$ are as before and, for $i < s$, $\rho_i$ ends with a robber's position. The elements $O_i$ are sets of vertices of $G$. The set $R_i$ represents the vertices occupied by robbers which are associated with $\rho_i$; $O_i$ is the set of vertices that history $\rho_i$ induces to be omitted while placing cops. To give a first idea of $O_i$, roughly, $O_i = \text{Reach}_{G-W_i}(R_i)$, but we shall see later that, in fact, the $O_i$ are more dynamic.

Now we drop the assumption that robbers from $R_i$ stay idle. They may prevent the cops to play as described up to now. One possibility is for one of them, say from $b_i \in R_i$, to jump to the robber $b_s$ of the longest history and then the robber from $b_i$ and from $b_s$ occupy vertices $b'_i$ and $b'_s$ that, after the cops' move, are in different SCCs of $C(b_s)$ (remember our definition of SSC). The cop player may not have additional cops to play in $C(b'_s)$. Thus the cop player has to reuse cops from team the team $U_i$ of robber who left $b_i$. However, the cops from $U_i$ cannot be just taken away before cop placements are made up that were omitted because the target vertices were reachable from $b_i$. Our solution is to let cops from team $U_i$ play according to $f$ from where they stopped. The cops' vertices are are stored in $\rho_i$; as the robber's vertices the vertex $b_{i+1}$ (of the next play) are taken. This is continued until the last position of the next stored play $\rho_{i+1}$ is reached by $\rho_i$. Then $\rho_i$ and $\rho_{i+1}$ are merged.

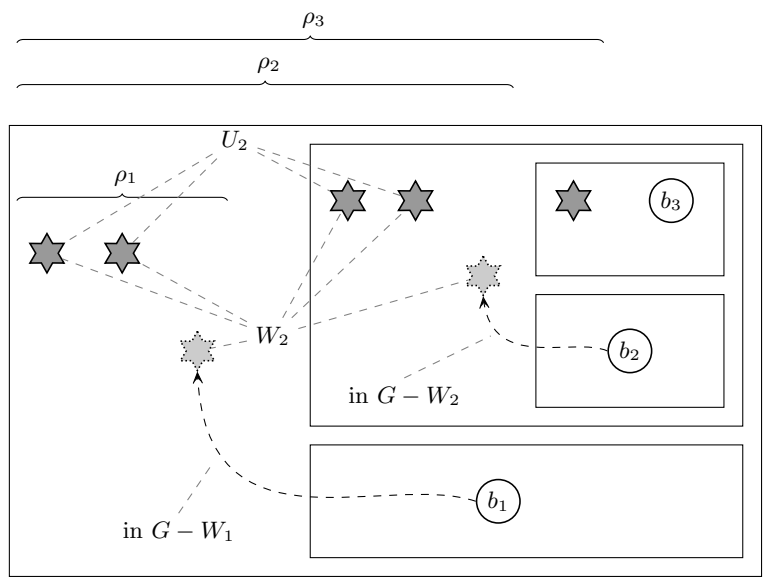

**Fig. 2.** Memory used by strategy $\otimes_r f$ and the graph $G$. Squares are robbers' components. Stars denote cops' vertices, dotted light gray stars denote vertices where cops placements were omitted.

The second case where the cops have to play in a different way is that the robber corresponding the longest history is caught or jumps away. In this case his SCC is not reachable for any robber any more, as the robbers play according to an isolating strategy. We take the cops from the graph placed since the last position in $\rho_{s-1}$, i.e., since the last time the robbers ran into different components. Then we choose another robber from $R_{s-1}$.

Now we present the strategy $\otimes_r f$ and the memory updates formally. We define the new set $U' = \otimes_r f((U, R), \zeta)$ of vertices occupied by cops and the new memory state

$$\zeta' = ((\rho'_1, R'_1, O'_1), \ldots, (\rho'_{s'-1}, R'_{s'-1}, O'_{s-1}), \rho'_{s'}).$$

We also maintain the following additional invariants. To describe them let

- $\text{last}(\rho_i) = (W_i^{-1}, W_i, b_i)$, for $i \in \{1, \ldots, s-1\}$,
- $\text{last}(\rho_s) \in \{(W_s, b_s), (W_s^{-1}, W_s, b_s)\}$,
- $U_i = W_i \setminus O^{i-1}$, $U^i = \bigcup_{j=1}^{i} U_j$ and $W^i = \bigcup_{j=1}^{i} W_j$ for $i \in \{1, \ldots, s\}$
- $R^i = \bigcup_{j=1}^{i} R_j$ and $O^i = \bigcup_{j=1}^{i} O_j$ for $i \in \{1, \ldots, s-1\}$.
- $R_s = \{b_s\}$, if $b_s \in R$ and $R_s = \emptyset$, else.

**Invariants and basic implications.**

(**Robs**) The sets $R_i$ are pairwise disjoint and $R = \bigcup_{i=1}^{s} R_i$.

(**Cops**) $U = \bigcup_{i=1}^{s} U_i$.

(**Omit**) For all $i \in \{1, \ldots, s-1\}$, $R_i \subseteq O_i = \text{Reach}_{G-W_i}(O_i)$.

(**Ext**) For all $i \in \{1, \ldots, s-1\}$, $O_i \subseteq \text{Reach}_{G-W_i^{-1}}(b_i)$.

Given the description above, (Robs) and (Cops) are the obvious formalizations of how the actual position in the game against $r$ robbers is connected to the several plays in the game against one robber that we maintain in the memory. Moreover, (Omit) and (Ext) formalize the important properties of the sets $O_i$ of positions where we have omitted placements of cops which we have also described above. The significance of this precise formulation will also become apparant in the following lemmata, which state several properties that can easily be derived from the invariants and which we will use frequently in the proof.

In addition to (Cops), we also assume that, if $(U, R)$ is a cops' position and $b_s \in R$ then $\text{last}(\rho_s) = (W_s, b_s)$.

The first part of (Omit) together with (Ext) guarantees that each robber that is associated with $\rho_i$ is also consistent with $\rho_i$.

**Lemma 11.** *For all $b \in R_i$, $\rho_i \cdot (W_i, b)$ is consistent with $f$.*

*Proof.* By (Omit) we have $b \in O_i$ and therefore, using (Ext), we obtain that $b$ is reachable from $b_i$ in $G - W_i^{-1}$. Moreover, as $\text{last}(\rho_i) = (W_i^{-1}, W_i, b_i)$ and $\rho_i$ is consistent with $f$ according to (Cons), $\rho_i \cdot (W_i, b)$ is consistent with $f$ as well. $\square$

**Lemma 12.**

(1) For any $i \in \{1, \ldots, s-1\}$ and any $b \in R_i$, $\text{Reach}_{G-W_i}(b) = \text{Reach}_{G-W^i}(b)$.
(2) $\text{Reach}_{G-W_s}(b_s) = \text{Reach}_{G-W^s}(b_s)$.

*Proof.* Consider some $i \in \{1, \ldots, s-1\}$ and some $b \in R_i$. As $W_i \subseteq W^i$, we have $\text{Reach}_{G-W_i}(b) \supseteq \text{Reach}_{G-W^i}(b)$, so assume that the converse inclusion $\text{Reach}_{G-W_i}(b) \subseteq \text{Reach}_{G-W^i}(b)$ does not hold. Then there is some $u \in W^{i-1} \setminus W_i$ such that $u \in \text{Reach}_{G-W_i}(b)$. Now if $j \in \{1, \ldots, i-1\}$ such that $u \in W_j$, then due to (Lin), $\rho_j \prec \rho_i$. Moreover, $\text{last}(\rho_j) = (W_j^{-1}, W_j, b_j)$ and, by Lemma 11, $\rho_i \cdot (W_i, b_i)$ is consistent with $f$, but as $\rho_j$ is consistent with $f$ as well due to (Cons), $\text{Reach}_{G-W_i}(b) \cap W_j \neq \emptyset$ contradicts the monotonicity of $f$ (which is violated in position $(W_i^{-1}, W_i, b_i)$). For $b_s$, the argument is the same. $\square$

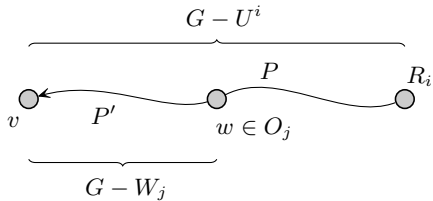

**Fig. 3.** $v \in \text{Reach}_{G-W_j}(O_j)$ implies $v \in O_j$ by (Omit)

The following lemma is one of the key arguments for monotonicity. It can be directly derived from (Omit) without using other invariants.

**Lemma 13.** $\text{Reach}_{G-U^i}(R_i) \subseteq O^i$.

*Proof.* Let $v \in \text{Reach}_{G-U^i}(R_i)$ and let $P$ be a path from $R_i$ to $v$ in $G - U^i$ as depicted in Figure 3. If $v \in \text{Reach}_{G-W_i}(R_i)$, by (Omit) we have $v \in \text{Reach}_{G-W_i}(O_i) = O_i \subseteq O^i$. Let therefore $v \notin \text{Reach}_{G-W_i}(R_i)$. Then $P \cap W_i \neq \emptyset$ and we consider the minimal $l \leq i$ such that $P \cap W_l \neq \emptyset$ and some $w \in P \cap W_l$. As $P \cap U^i = \emptyset$ we have $w \notin U^i \supseteq U_l$ and by definition of $U_l$ this yields $w \in O^{l-1}$, that means, $w \in O_j$ for some $j < l$. Now $v$ is reachable from $w$ in $G$ via some path $P' \subseteq P$ and, due to the minimal choice of $l$, $P \cap W_j = \emptyset$. Hence, $P' \cap W_j = \emptyset$, see Figure 3. This yields $v \in \text{Reach}_{G-W_j}(w) \subseteq \text{Reach}_{G-W_j}(O_j)$ and as, by (Omit), $\text{Reach}_{G-W_j}(O_j) = O_j$ it follows that $v \in O_j \subseteq O^i$. □

Finally, we formulate the fact that the reachability area of a robber is not restricted by cops of longer histories as a direct corollary of Lemma 13.

**Corollary 14.** *For all $i \in \{1, \ldots, s-1\}$ and all $b \in R_i$ we have $\text{Reach}_{G-U}(b) = \text{Reach}_{G-U^i}(b)$.*

**Initial Move.** As we assumed that $G$ is strongly connected, by Lemma 8, the robbers do not split in the first move. So let the initial move be $\bot \to (\emptyset, \{b\})$. After the move, the memory state is set to $((\emptyset, b))$. All the invariants hold obviously for $(\emptyset, \{b\})$ and $((\emptyset, b))$.

Now consider some game position $(U, R)$ where it is the cops player's turn and some memory state $\zeta$ such that all invariants are fulfilled.

**Move of the Cops.** In the following, we define the new set $U' = \otimes_r f((U, R), \zeta)$ of vertices occupied by cops and the new memory state

$$\zeta' = ((\rho'_1, R'_1, O'_1), \ldots, (\rho'_{s'-1}, R'_{s'-1}, O'_{s-1}), \rho'_{s'}).$$

*Case I: $b_s \notin R$*
That means, the robber $b_s$ which is stored in the longest history is not on the graph anymore. Hence, if $s = 1$ then this robber has been caught and as there are no other robbers, all the robbers are caught and the cops have won. Otherwise, we set $U' := U^{s-1} = \bigcup_{i=1}^{s-1} U_i$, that means we remove the cops corresponding to the longest history from the graph. For the memory update, we consider the longest prefix $\rho_{s-1}$ of $\rho_s$ that we have maintained and we distinguish two cases:

- If $R_{s-1} = \emptyset$:
  The new memory state $\zeta'$ is obtained from $\zeta$ by deleting $\rho_s$ and replacing $(\rho_{s-1}, R_{s-1}, O_{s-1})$ by the history $\rho_{s-1} \cdot (W_{s-1}, b_s)$.
- If $R_{s-1} \neq \emptyset$:
  In this case we have to select one of the robbers from $R_{s-1}$ that we want to pursue next. Choose some robber $b \in R_{s-1}$ and define the new set $\tilde{O}_{s-1} := \text{Reach}_{G-W_{s-1}}(R_{s-1} \setminus \{b\})$. Then the new memory state $\zeta'$ is obtained from $\zeta$ by replacing $(\rho_{s-1}, R_{s-1}, O_{s-1})$ by $(\rho_{s-1}, R_{s-1} \setminus \{b\}, \tilde{O}_{s-1})$ and replacing $\rho_s$ by $\rho_{s-1} \cdot (W_{s-1}, b)$.

*Case II:* $b_s \in R$.

*Case II.1:* There is some $i \in \{1, \ldots, s-1\}$ such that $R_i = \emptyset$.
That means, there is no robber associated with history $i$. First, consider the next robbers' move according to $\rho_{i+1}$. (Note that $i < s$.) That is, consider $\tilde{b}_i \in V$ and some suffix $\eta$ of $\rho_{i+1}$ such that $\rho_{i+1} = \rho_i(W_i, \tilde{b}_i)\eta$. Now we distinguish three more cases.

(a) $\rho_{i+1} = \rho_i \cdot (W_i, \tilde{b}_i) = \rho_s$, i.e., $\eta$ is empty.
    Set $U' := U$, and update the memory by deleting $(\rho_i, R_i, O_i)$ from $\zeta$.

For the other cases, we set

- $\tilde{W}_i := f(W_i, \tilde{b}_i)$ and
- $U' := \bigcup_{j \neq i} U_j \cup (\tilde{W}_i \setminus O^{i-1})$.
- $\tilde{O}_i = (O_i \cap \mathrm{Reach}_{G-W_i}(\tilde{b}_i)) \setminus \tilde{W}_i$ and
- $\tilde{\rho}_i = \rho_i \cdot (W_i, \tilde{b}_i) \cdot (W_i, \tilde{W}_i, \tilde{b}_i)$

(b) $\tilde{\rho}_i \neq \rho_{i+1}$.
    That means, we have not reached the end of the next history. In this case, we replace $(\rho_i, R_i, O_i)$ by $(\tilde{\rho}_i, R_i, \tilde{O}_i)$.
(c) $\tilde{\rho}_i = \rho_{i+1}$.
    The memory update is to replace $(\rho_{i+1}, R_{i+1}, O_{i+1})$ by $(\rho_{i+1}, R_{i+1}, O_{i+1} \cup \tilde{O}_i)$ and to remove $(\rho_i, R_i, O_i)$.

*Case II.2:* For all $i \in \{1, \ldots, s-1\}$ we have $R_i \neq \emptyset$.
In this case the cops play against the robber from $\rho_s$. We define

- $\tilde{W}_s = f(W_s, b_s)$ and
- $U' := \bigcup_{j < s} U_j \cup (\tilde{W}_s \setminus O^{s-1})$.

and for the memory update, we replace $\rho_s$ by $\rho'_s = \rho_s \cdot (W_s, \tilde{W}_s, b_s)$.

Now we prove that the move of the cops from $U$ to $U'$ is monotone, that means, no robber can reach any vertex due to this move which was previously blocked for all robbers. As the cops from $U \cap U'$ are precisely those which remain idle this means that no robber can reach any vertex from $U \setminus U'$ in $G - (U \cap U')$.

**Lemma 15.** $(U \setminus U') \cap \mathrm{Reach}_{G-(U \cap U')}(R) = \emptyset$.

*Proof.* If $b_s \notin R$ (Case I) we have $U' = U^{s-1}$ so, by (Cops), $U^{s-1} \subseteq U \cap U'$. Moreover, (Robs) yields $R = \bigcup_{i=1}^{s-1} R_i$ and hence, using Lemma 13, we obtain $\mathrm{Reach}_{G-(U \cap U')}(R) \subseteq O^{s-1}$. Now due to the definition of $U_s$ we have $O^{s-1} \cap U_s = \emptyset$, which yields $\mathrm{Reach}_{G-(U \cap U')}(R) \cap U_s = \emptyset$ and thus, by (Cops), the move of $\otimes_r f$ is monotone in this case.

Now assume that $b_s \in R$ (Case II) and first consider Case II.1, i.e., there is some $i \in \{1, \ldots, s-1\}$ such that $R_i = \emptyset$. In Subcase (a) the cops don't move, so the move is monotone. Otherwise we have $U' = \bigcup_{j \neq i} U_j \cup \tilde{U}_i$ with $\tilde{U}_i = \tilde{W}_i \setminus O^{i-1}$ where $\tilde{W}_i = f(W_i, \tilde{b}_i)$ and $\rho_{i+1} = \rho_i(W_i, \tilde{b}_i)\eta$ are as above. So we have extended history $\rho_i$ by the next move of the robber according to $\rho_{i+1}$ and the next move of the cops according to $f$, omitting the vertices from $O^{i-1}$. Now assume that this move is not monotone, i.e., there is some $v \in U \setminus U'$ with $v \in \text{Reach}_{G-(U \cap U')}(R)$. By definition of $U'$ and (Cops), $v \in U_i \setminus \tilde{U}_i$.

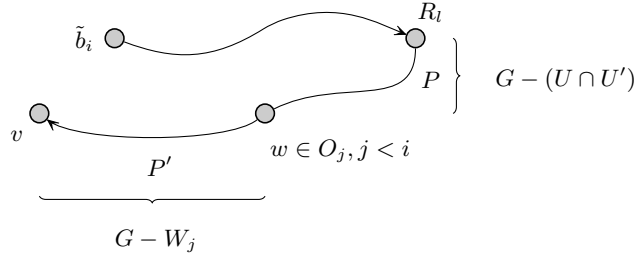

**Fig. 4.** Robbers from longer histories than $\rho_i$ cannot cause non-monotonicity

We distinguish, which robbers can reach $v$. First, consider robbers from smaller histories than $\rho_i$, that means, from the set $R^{i-1}$. As $U^{i-1} \subseteq U \cap U'$, using Lemma 13 we obtain $\text{Reach}_{G-(U \cap U')}(R^{i-1}) \subseteq O^{i-1}$. Since, due to the definition of $U_i$, $O^{i-1} \cap U_i = \emptyset$ we have $v \notin \text{Reach}_{G-(U \cap U')}(R^{i-1})$, that means, no robbers from $R^{i-1}$ can cause non-monotonicity. So, as $R_i = \emptyset$, we have $v \in \text{Reach}_{G-(U \cap U')}(R^{>i})$ where $R^{>i} = \bigcup_{l=i+1}^{s-1} R_l \cup \{b_s\}$ denotes the set of robbers from longer histories than $\rho_i$. Consider some path $P$ from $R^{>i}$ to $v$ in $G - (U \cap U')$ as depicted in Figure 4.

First, we show that $v \notin \text{Reach}_{G-(W_i \cap \tilde{W}_i)}(R^{>i})$. For $l \in \{i+1, \ldots, s-1\}$ and any $b \in R_l$, by (Lin), $\rho_i \cdot (W_i, \tilde{b}_i)$ is a strict prefix of $\rho_l \cdot (W_l, b)$ and, by Lemma 11, both these histories are consistent with $f$. So, by monotonicity of $f$, any robber $b \in R_l$ is reachable from $\tilde{b}_i$ in $G - W_i$ and hence in $G - (W_i \cap \tilde{W}_i)$. Moreover, as we are in Case II.1 (b) or (c), the same reasons can be used to show that $b_s$ is also reachable from $\tilde{b}_i$ in $G - W_i$ and hence in $G - (W_i \cap \tilde{W}_i)$. Therefore, if $v \in \text{Reach}_{G-(W_i \cap \tilde{W}_i)}(R^{>i})$ then $v \in \text{Reach}_{G-(W_i \cap \tilde{W}_i)}(\tilde{b}_i)$. But as $v \in U_i \subseteq W_i$ this contradicts monotonicity of $f$ since $\rho_i \cdot (W_i, \tilde{b}_i) \cdot (W_i, \tilde{W}_i, \tilde{b}_i)$ is consistent with $f$. Hence, $v \notin \text{Reach}_{G-(W_i \cap \tilde{W}_i)}(R^{>i})$.

So $P$ is a path from $R^{>i}$ to $v$ in $G - (U \cap U') = \emptyset$ and as there is no such path in $G - (W_i \cap \tilde{W}_i)$ we conclude $P \cap (W_i \cap \tilde{W}_i) \neq \emptyset$. We consider the minimal $l \leq i$ such that $P \cap \widehat{W}_l \neq \emptyset$ where we set $\widehat{W}_j = W_j$ for $j < i$ and $\widehat{W}_i = W_i \cap \tilde{W}_i$, analogously for $\widehat{U}_j$. That means, $\widehat{W}_j$ are vertices occupied by cops according to $\rho_j$ which remained idle in the last move. Now let $w \in P \cap \widehat{W}_l$. First, as $w \in P$, $w \notin U \cap U'$ so (Cops) and the definition of $U'$ yield $w \notin \widehat{U}_l$. Therefore, $w \in \widehat{W}_l \setminus \widehat{U}_l$ and hence, using the definition of $U_l$, and of $\tilde{U}_i$ if $l = i$, we obtain $w \in O^{l-1}$, that means, $w \in O_j$ for some $j < l$. Moreover, $v$ is reachable from $w$ in $G$ via some path $P' \subseteq P$ and, due to the minimal choice of $l$, $P' \cap \widehat{W}_j = P' \cap W_j = \emptyset$, so $v \in \text{Reach}_{G-W_j}(w) \subseteq \text{Reach}_{G-W_j}(O_j) = O_j \subseteq O^{i-1}$. The last equality is due to (Omit). But as $O^{i-1} \cap U_i = \emptyset$, $v \in O^{i-1}$ is a contradiction to $v \in U_i$.

Finally, consider Case II.2, i.e., for all $i \in \{1, \ldots, s-1\}$ we have $R_i \neq \emptyset$. First notice that, due to definition of $U'$ and invariant (Cops), $U \setminus U' \subseteq U_s$. For robbers other than $b_s$ the same arguments as in Case I and Case II.1, using (Robs) and Lemma 13, show that they cannot cause non-monotonicity. The argument for $b_s$ is the same as in Case II.1: assume that $b_s$ causes non-monotonicity at some vertex $v$. As $\rho_s$ is consistent with $f$ due to (Cons) and $f$ is monotone, $b_s$ can reach $v$ only via $O^{s-1}$ (using (Cops)). However, as usual, $O^{s-1}$ is a trap for the robbers in $G - U$ and $v$ cannot be in $O^{s-1}$, so this is impossible. □

Now we prove that after the move of the cops, the invariants that we have formulated still hold. Note that for $(U, R)$ and $\zeta$ we have assumed that all the invariants hold by induction. We first give a separate lemma for (Robs), (Lin), (Cons) and (Ext) and we prove them quite briefly as they can be obtained easily from the induction hypothesis, using the definition of the cops' move.

**Lemma 16.** *(Robs), (Lin), (Cons) and (Ext) are preserved by the cops' move.*

*Proof.* (Robs) follows immediately from the induction hypothesis. Moreover, linearity of $\prec$ is obviously preserved in Case I, Case II.1 (a) and (b) and in Case II.2 of the cops' move. In Case II.1 (b) we have to show that $\tilde{\rho}_i \prec \rho_{i+1}$. First notice that $\rho_i \cdot (W_i, \tilde{b}_i) \prec \rho_{i+1}$ as $\rho_{i+1} = \rho_i \cdot (W_i, \tilde{b}_i)\eta$ and $\eta \neq \emptyset$. Further, the first position in $\eta$ is $(W_i, \tilde{W}_i, \tilde{b}_i)$ as $\rho_{i+1}$ is consistent with $f$ according to (Cons) and $\tilde{W}_i = f(W_i, \tilde{b}_i)$. As $\tilde{\rho}_i \neq \rho_{i+1}$ it follows that $\tilde{\rho}_i \prec \rho_{i+1}$. This establishes (Lin).

For (Cons), consider first Case I. If $R_{s-1} = \emptyset$ then $\rho'_{s'} = \rho_{s-1} \cdot (W_{s-1}, b_s)$. As $\text{last}(\rho_s) \in \{(W_s^{-1}, W_s, b_s), (W_s, b_s)\}$ and $\rho_{s-1} \prec \rho_s$ and due to (Cons) both these histories are consistent with $f$, which is monotone, $b_s$ is reachable from $b_{s-1}$ in $G - (W_{s-1}^{-1} \cap W_{s-1})$, so $\rho_{s-1} \cdot (W_{s-1}, b_s)$ is consistent with $f$. If $R_{s-1} \neq \emptyset$ then $\rho_{s-1} \cdot (W_{s-1}, b)$ is consistent with $f$ for any $b \in R_{s-1}$ due to Lemma 11. In Case II.1 (a) and (b), (Cons) follows immediately from the induction hypothesis. In Case II.1 (b), (Cons) follows from (Lin) as $\rho'_{s'} = \rho_s$ is consistent with $f$ and $\tilde{\rho}_i \prec \rho_s$. Finally, in Case II.2, $\rho_s$ is consistent with $f$ due to (Cons) and $W'_s = f(W_s, b_s)$, so $\rho'_{s'} = \rho'_s$ is consistent with $f$ as well.

To prove (Ext) first notice that in Case I, if $R_{s-1} = \emptyset$ (Ext) follows immediately from the induction hypothesis. Moreover, if $R_{s-1} \neq \emptyset$ then we have $s' = s$ and we have to show that $O'_{s-1} = \tilde{O}_{s-1} \subseteq \text{Reach}_{G-W_{s-1}^{-1}}(b_{s-1})$. As, by Lemma 11, for any $b' \in R_{s-1}$ the history $\rho_{s-1}(W_{s-1}, b')$ is consistent with $f$ which is monotone, the reachability area of any $b' \in R_{s-1}$ in $G - W_{s-1}$ is a subset of the reachability area of $b_{s-1}$ in $G - W_{s-1}^{-1}$. Hence, by definition of $\tilde{O}_{s-1}$, the statement follows. In Case II, (Ext) follows easily from the induction hypothesis, using the definition of $\tilde{O}_i$ in Case II.1 (b) and (c). □

For the remaining two invariants (Omit) and (Cops) we have two separate lemmata which we prove in greater detail. The most interesting cases in the proofs of these two invariants are Case II.1 (b) and (c). The crucial point here is the new set $\tilde{O}_i$. See Figure 5 for an illustration.

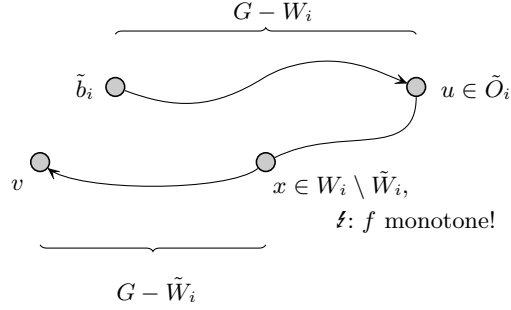

**Fig. 5.** $\tilde{O}_i$ is closed under reachability in $G - \tilde{W}_i$

**Lemma 17.** *(Omit) is preserved by the cops' move.*

*Proof.* In Case I, if $R_{s-1} = \emptyset$, (Omit) follows immediately from the induction hypothesis, so consider the case where $R_{s-1} \neq \emptyset$. Then $s' = s$, $W'_{s-1} = W_{s-1}$ and $O'_{s-1} = \tilde{O}_{s-1} = \text{Reach}_{G-W_{s-1}}(R_{s-1} \setminus \{b\})$. Clearly, this yields that $O'_{s-1}$ is closed under reachability in $G - W_{s-1}$. Moreover, by (Omit), $R_{s-1} \subseteq \text{Reach}_{G-W_{s-1}}(O_{s-1})$ so we have $R_{s-1} \cap W_{s-1} = \emptyset$ an hence $R_{s-1} \setminus \{b\} \subseteq \text{Reach}_{G-W_{s-1}}(R_{s-1} \setminus \{b\}) = O'_{s-1}$.

Now consider Case II.1. In Case (a), (Omit) follows immediately from the induction hypothesis. In Case (b), $R'_i \subseteq O'_i$ is trivial as $R'_i = R_i = \emptyset$, so we have to show that $O'_i = \text{Reach}_{G-W'_i}(O'_i)$. First, notice that $O'_i = \tilde{O}_i = (O_i \cap \text{Reach}_{G-W_i}(\tilde{b}_i)) \setminus \tilde{W}_i$ and $W'_i = \tilde{W}_i = f(W_i, \tilde{b}_i)$. Moreover, by definition of $\tilde{O}_i$ in this case we have $\tilde{O}_i \cap \tilde{W}_i = \emptyset$, so $\tilde{O}_i \subseteq \text{Reach}_{G-\tilde{W}_i}(\tilde{O}_i)$.

Now we show that $\tilde{O}_i$ is closed under reachability in $G - \tilde{W}_i$. Let $v \in \text{Reach}_{G-\tilde{W}_i}(\tilde{O}_i)$. Clearly, $v \notin \tilde{W}_i$. Now let $u \in \tilde{O}_i$ such that $v$ is reachable from $u$ in $G - \tilde{W}_i$. As $\tilde{O}_i = (O_i \cap \text{Reach}_{G-W_i}(\tilde{b}_i)) \setminus W_i$ we have $u \in \text{Reach}_{G-W_i}(\tilde{b}_i)$ and $v \in \text{Reach}_{G-\tilde{W}_i}(u)$. Therefore, there is a cop-free path from $\tilde{b}_i$ to $v$ via $u$ in $G - (W_i \cap \tilde{W}_i)$. By (Cons), $\rho_i$ is consistent with $f$ and $\tilde{W}_i = f(W_i, \tilde{b}_i)$, so, as $f$ is monotone, this path must be cop-free in $G - W_i$, see Figure 5. Thus, $v \in \text{Reach}_{G-W_i}(u)$ and as $u \in O_i$ (by definition of $\tilde{O}_i$) and $u \in \text{Reach}_{G-W_i}(\tilde{b}_i)$ we have $v \in \text{Reach}_{G-W_i}(O_i)$ and $v \in \text{Reach}_{G-W_i}(\tilde{b}_i)$. Moreover, by invariant (Omit) we have $\text{Reach}_{G-W_i}(O_i) = O_i$, so $v \in O_i \cap \text{Reach}_{G-W_i}(\tilde{b}_i)$ and as $v \notin \tilde{W}_i$ this yields $v \in \tilde{O}_i$.

In Case (c), we have to show that $R'_i \subseteq O'_i$ and that $O'_i$ is closed under reachability in $G - W'_i$. First notice that $R'_i = R_{i+1}$, $O'_i = O_{i+1} \cup \tilde{O}_i$ and $W'_i = W_{i+1}$. Now, by (Omit), $R_{i+1} \subseteq O_{i+1} \subseteq O_{i+1} \cup \tilde{O}_i$. Moreover, as in Case (b), $\tilde{O}_i$ is closed under reachability in $G - \tilde{W}_i$ and as $\tilde{\rho}_i = \rho_{i+1}$ we have $\tilde{W}_i = W_{i+1}$. By (Omit), $O_{i+1}$ is closed under reachability in $G - W_{i+1}$, so the union $O_{i+1} \cup \tilde{O}_i$ is closed under reachability in $G - W_{i+1}$ as well. Finally, in Case II.2, (Omit) follows again from the induction hypothesis. □

**Lemma 18.** *(Cops) is preserved by the cops' move.*

*Proof.* We have to show that $U' = \bigcup_{j=1}^{s'} U'_j$ where $s' \in \{s-1, s\}$ is the length of $\zeta'$. Note that, by definition, $U'_j = W'_j \setminus (O^{j-1})'$ for $j = 1, \ldots, s'$.

In Case I, Case II.1 (a) and Case II.2, this can easily be obtained using the induction hypothesis and the definition of $U'$. Now consider Case II (b). Then $s' = s$ and $O'_j = O_j$ for $j \neq i$ and $O'_i = \tilde{O}_i \subseteq O_i$. As, moreover, $W'_j = W_j$ for $j < i$, we have $U'_j = U_j$ for $j < i$. Furthermore, $U'_i = W'_i \setminus (O^{i-1})' = \tilde{W}_i \setminus O^{i-1}$ and as $(O^{j-1})' \subseteq O^{j-1}$ for $j = 1, \ldots, s$ we have $U_j \subseteq U'_j$ for $j > i$. Hence, $U' \subseteq \bigcup_{j=1}^{s} U'_j$ and it remains to show $\bigcup_{j=1}^{s} U'_j \subseteq U'$.

So assume that there is some $v \in (\bigcup_{j=1}^{s} U'_j) \setminus U'$. Then $v \in U'_j$ for some $j > i$ and as $v \notin U' \supseteq U_j$ we have $v \in W_j \setminus (O^{j-1})'$ but $v \notin O^{j-1}$. Moreover, since $O'_l = O_l$ for $l \neq i$ we have $v \in O_i \setminus O'_i = O_i \setminus \tilde{O}_i$. So, by definition of $\tilde{O}_i$, we have $v \in \tilde{W}_i$ or $v \notin \text{Reach}_{G-W_i}(\tilde{b}_i)$. As $v \notin U'$ we have $v \notin \tilde{W}_i \setminus O^{i-1}$ and as $v \notin O^{j-1} \supseteq O^{i-1}$ it follows that $v \notin \tilde{W}_i$, so $v \notin \text{Reach}_{G-W_i}(\tilde{b}_i)$. Now let $\rho = \hat{\rho}(W^{-1}, W, b)$ be the shortest prefix of $\rho_j$ such that $v \in W$. Note that such a prefix exists as $v \in W_j$. Now due to (Cons), $\tilde{\rho}_i$ and $\rho$ are consistent with $f$ and $f$ is monotone, so since $v \notin \tilde{W}_i$ we have $\tilde{\rho}_i \prec \rho$ and as $v \notin \text{Reach}_{G-W_i}(\tilde{b}_i)$ we also have $v \notin \text{Reach}_{G-W^{-1}}(b)$. But this is a contradiction to the fact that $f$ does not place cops on vertices which are already unavailable for the robber.

Finally, in Case (c), we have $s' = s - 1$ as we delete the $i$-th element of $\zeta$. Hence, we have a shift of indices. Accounting for this fact, (Cops) can be proved with very similar arguments as in Case (b). □

**Move of the Robbers.** Let $R'$ be the set of vertices occupied by robbers after their move. If $R' = R$ we do not update the memory. This happens in particular after the cops' move in Case I and in Case II.1 (a) of the cops' move: there, we do not place cops on the graph and as the robbers use a prudent strategy, $R' = R$. We shall not consider these cases.

Assume that $R' \neq R$ and consider the memory state

$$\zeta = ((\rho_1, R_1, O_1), \ldots, (\rho_{s-1}, R_{s-1}, O_{s-1}), \rho_s)$$

before the robbers' move from $R$ to $R'$. Note that, in particular, we have $b_s \in R$. We assign every robber from $R'$ to some unique history $\rho_i$, $i = 1, \ldots, s$ which yields, for any $i \in \{1, \ldots, s\}$ the new set $\tilde{R}_i$. So consider any robber $b \in R'$. If $b \in O^{s-1}$ then let $i = \min\{j \in \{1, \ldots, s-1\} \mid b \in O_j\}$ and assign $b$ to $\rho_i$. Otherwise assign $b$ to $\rho_s$.

In the following, we also need the memory state

$$\overline{\zeta} = ((\overline{\rho_1}, \overline{R}_1, \overline{O}_1), \ldots, (\overline{\rho}_{\overline{s}-1}, \overline{R}_{\overline{s}-1}, \overline{O}_{s-1}), \overline{\rho}_{\overline{s}})$$

and the set $U^{-1}$ of vertices occupied by cops, before the move of the cops.

The crucial point which we have to prove about the memory update after the move of the robbers is that our assignment of robbers to histories is meaningful in the sense that a robber which has been assigned to a certain history is also consistent with this history according to $f$. For the robbers assigned to histories $\rho_i$ with $i < s$ this follows easily from the fact that $\tilde{R}_i \subseteq O_i$, similar as in Lemma 11. For the robbers in $\tilde{R}_s$ this is, however, much more involved. We have to show that each such robbers can be reached from $\overline{b}_{\overline{s}} = b_s$ in the graph $G - \overline{W}_{\overline{s}}$ which then shows that prolonging the longest history by a move from $b_s$ to some robber from $\tilde{R}_s$ yields again an $f$-history. This property is proved in the following lemma.

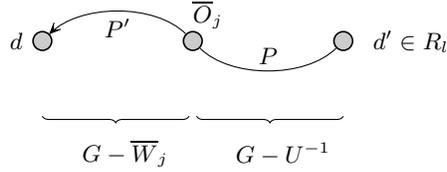

**Fig. 6.** Any $d \in \tilde{R}_s \setminus \text{Reach}_{G-\overline{W}_{\overline{s}}}(\overline{b}_{\overline{s}})$ is in $\overline{O}^{s-1}$.

**Lemma 19.** $\tilde{R}_s \subseteq \text{Reach}_{G-\overline{W}_{\overline{s}}}(\overline{b}_{\overline{s}})$.

*Proof.* Let $d \in \tilde{R}_s$. As the robbers have moved from $R$ to $R'$ in their move, there is some $d' \in R$ such that $d$ is reachable from $d'$ in $G - (U^{-1} \cap U)$. As we have already shown in Lemma 15, the move from $U^{-1}$ to $U$ was monotone, so $d$ is reachable from $d'$ in $G - U^{-1}$. Let $P$ be a path from $d'$ to $d$ in $G - U^{-1}$ and assume that $d \notin \text{Reach}_{G-\overline{W}_{\overline{s}}}(\overline{b}_{\overline{s}})$. We show that then $d \in O^{s-1}$ in contradiction to $d \in \tilde{R}_s$ as by definition of $\tilde{R}_s$, $\tilde{R}_s \cap O^{s-1} = \emptyset$. By (Robs) for $\overline{\zeta}$, $R = \bigcup_{i=1}^{\overline{s}}(\overline{R}_i)$, so there is some (unique) $l \leq \overline{s}$ with $d' \in \overline{R}_l$.

First we show $d \in \overline{O}^{\overline{s}-1}$, see Figure 6. If $d' \neq \overline{b}_{\overline{s}}$ then according to (Omit) for $\overline{\zeta}$ we have $d' \in \overline{R}_l \subseteq \overline{O}_l$ and as $d' \in P$, we have $P \cap \overline{O}^l \neq \emptyset$. In the other case we have $d' = \overline{b}_{\overline{s}}$ so $d \in \text{Reach}_{G-U^{-1}}(\overline{b}_{\overline{s}})$ and as, by (Cops) for $\overline{\zeta}$, $\overline{U}_{\overline{s}} \subseteq U^{-1}$ we have $d \in \text{Reach}_{G-\overline{U}_{\overline{s}}}(\overline{b}_{\overline{s}})$. However, by our assumption, $d \notin \text{Reach}_{G-\overline{W}_{\overline{s}}}(\overline{b}_{\overline{s}})$, so by definition of $\overline{U}_{\overline{s}}$, $P \cap \overline{O}^{\overline{s}-1} \neq \emptyset$. Hence, in any case we have $P \cap \overline{O}_j \neq \emptyset$ for some $j \leq \min\{\overline{s}-1, l\}$ and we consider the minimal such $j$. Then, by (Cops) for $\overline{\zeta}$, $\overline{U}_j \subseteq U^{-1}$, so $d$ is reachable from $\overline{O}_j$ in $G - \overline{U}_j$ via a path $P' \subseteq P$, see Figure 6. So if $d \notin \text{Reach}_{G-\overline{W}_j}(\overline{O}_j)$ then by definition of $\overline{U}_j$ we have $P \cap \overline{O}^{j-1} \neq \emptyset$ which contradicts minimality of $j$. Hence, $d \in \text{Reach}_{G-\overline{W}_j}(\overline{O}_j) = \overline{O}_j$ by (Omit) for $\overline{\zeta}$.

Now we show that $d$ is also in $O^{s-1}$. We distinguish the moves that the cops may have made. Case I and Case II.1 (a) of the cops' move do not have to be considered here as discussed above. If $\overline{O}_j = O_j$, which in particular holds in Case II.2, then $d \in O_j \subseteq O^{s-1}$. Now assume that $\overline{O}_j \neq O_j$, so we are in Case II.1 (b) or (c). Let $i$ be as in these cases. Then for all $m < i$, we have $\overline{O}_m = O_m$, so $j \geq i$. Morover, for all $m > i$, $\overline{O}_m = O_m$ (in Case II.(b)) or $\overline{O}_m \subseteq O_{m-1}$ (in

Case II.(c)), so either $d \in O^{s-1}$ or $j \leq i$. The remaining case is $j = i$. Note that in this case, $j < l$ as either $l = \bar{s}$ and $j \leq \bar{s} - 1$ or $l < \bar{s}$. In the latter case, the reason is that $j \leq l$ and $\overline{R}_j = \overline{R}_i = \emptyset$ and $d' \in \overline{R}_l \neq \emptyset$. We show that $d \in \tilde{O}_j$, then by definition of the memory update $d \in O_j$ and hence $d \in O^{s-1}$.

By definition, $\tilde{O}_j = (\overline{O}_j \cap \text{Reach}_{G-\overline{W}_j}(\tilde{b}_j)) \setminus \tilde{W}_j$ where $\tilde{W}_j = W'_j = W_j$ and $\tilde{b}_j = b_j$. We have already shown that $d \in \overline{O}_j$. In order to see that $d \notin W_j$ notice that $d \in R'$, and $U_j \subseteq U$ according to (Cops), so $d \notin U_j$. Hence, if $d \in W_j$, we have $d \in \overline{O}^{j-1} = O^{j-1}$ by definition of $U_j$, contradicting $d \in \tilde{R}_s$. Thus, $d \notin W_j$ and it remains to show that $d \in \text{Reach}_{G-\overline{W}_j}(b_j)$. First notice that since $j < l$, we have $\tilde{\rho}_j \preccurlyeq \overline{\rho}_{j+1} \preccurlyeq \overline{\rho}_l$. So as, according to (Cons), all these histories are consistent with $f$, which is monotone, $\overline{b}_l$ is reachable from $b_j$ in $G - \overline{W}_j$, see Figure 7. Now if $l < \bar{s}$, $d' \in \overline{R}_l$, so by (Ext), $d'$ is reachable from $\overline{b}_l$ in $G - \overline{W}_l^{-1}$. Moreover, using again that $\tilde{\rho}_j \preccurlyeq \overline{\rho}_l$ are both consistent with $f$ and that $f$ is monotone, this yields that $d'$ is reachable from $\overline{b}_l$ in $G - \overline{W}_j$. If, on the other hand, $l = \bar{s}$ then $d' = \overline{b}_{\bar{s}} = \overline{b}_l$, so clearly, $d'$ is reachable from $\overline{b}_l$ in the graph $G - \overline{W}_j$. Therefore, $d'$ is reachable from $b_j$ in the graph $G - \overline{W}_j$ and as, by (Cops), $\overline{U}_j \subseteq U^{-1}$, $d$ is reachable from $d'$ in $G - \overline{U}_j$ via $P$. Hence, if $d$ is not reachable from $b_j$ in $G - \overline{W}_j$, then due to the definition of $\overline{U}_j$ there is some vertex from $\overline{O}^{j-1}$ on the path $P$ which contradicts the minimality of $j$. Hence, $d \in \text{Reach}_{G-\overline{W}_j}(b_j)$. □

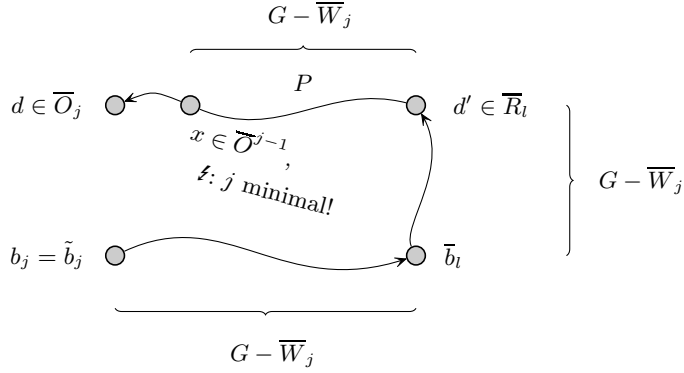

**Fig. 7.** If $i = j < l$, the robber $\tilde{b}_j$ can still reach $d$ in the graph $G - \overline{W}_j$ via $d'$.

Now we define the update of the memory from $\zeta$ to the new state
$$\zeta' = ((\rho'_1, R'_1, O'_1), \ldots, (\rho'_{s'-1}, R'_{s'-1}, O'_{s'-1}), \rho'_{s'}).$$
For the update we distinguish three cases according to the number of robbers that have been assigned to $\tilde{R}_s$, that means, which cannot be associated with any history $\rho_i$ for $i \in \{1, \ldots, s-1\}$, and according to whether the last position of $\rho_s$ is a cops' or a robbers' position. First, to simplify the case distinction, we prove that if we did not play against the robber from the longest history in the last move of the cops, then at most the robber $b_s = \bar{b}_{\bar{s}}$ can be consistently associated with $\rho_s$. For this, we need the assumption that the robbers use a prudent strategy.

**Lemma 20.** *If $\rho_s$ ends with a position of the cop player then $\tilde{R}_s \subseteq \{b_s\}$.*

*Proof.* Assume that $\rho_s$ ends with a cops' position, that means, $\rho_s = \hat{\rho}_s(W_s, b_s)$. So the last move of the cops was not as in Case II.2 and hence, as Case I and Case II.1 (a) do not have to be considered as discussed above, we have $W_s = \overline{W_{\bar{s}}}$ (and $b_s = \overline{b_{\bar{s}}}$). So Lemma 19 yields $\tilde{R}_s \subseteq \text{Reach}_{G-W_s}(b_s)$. By Lemma 12 we have $\text{Reach}_{G-W_s}(b_s) = \text{Reach}_{G-W^s}(b_s)$, so $\tilde{R}_s \subseteq \text{Reach}_{G-W^s}(b_s) \subseteq \text{Reach}_{G-U}(b_s)$. Therefore, since $b_s \in R$, $\tilde{R}_s \not\subseteq \{b_s\}$ contradicts the assumption that the robbers use a prudent strategy. □

Due to this property, we now need to distinguish only two other cases. Either, the last move of the cops was to play against the robber from the longest history $b_s$ and we have $|\tilde{R}_s| \geq 1$, that means, at least one of the robbers from $R'$ can still be consistently associated with the corresponding history $\rho_s$. (That all robbers from $\tilde{R}_s$ can be consistently associated with $\rho_s$ follows from (Cons) which will be proved in Lemma 21.) Otherwise, either we have not played against the robber from the longest history or we have played against the robber from the longest history and he has either been caught or he has returned to an earlier situation.
*Case 1:* $\rho_s$ ends with a position of the robber player and $|\tilde{R}_s| \geq 1$.

We choose one of the robbers $b \in \tilde{R}_s$ which we pursue further (that means, $b$ will be the new robber from the longest history), and we add a new history $\rho_{s'} = \rho_{s+1}$ which extends the history $\rho_s$ by the robbers' from $b_s$ to $b$. Moreover, we associate with the history $\rho_s$ the remaing robbers $\tilde{R}_s \setminus \{b\}$. Consequently, we also have to define a new set $O'_{s'-1} = O'_s$ which contains exactly the vertices reachable from the robbers in $\tilde{R}_s \setminus \{b\}$ in the graph $G - W_s$. Note that if $|\tilde{R}_s| = 1$ then $\tilde{R}_s \setminus \{b\} = \emptyset$ and $O'_s = \emptyset$, so in fact, $\rho_s$ does not have to be maintained anymore. However, this will be taken care of during the next move of the cops. Summarizing, we choose some $b \in \tilde{R}_s$ define $\tilde{O}_s = \text{Reach}_{G-W_s}(\tilde{R}_s \setminus \{b\})$ and set

$$\zeta' = \big((\rho_1, \tilde{R}_1, O_1), \ldots, (\rho_{s-1}, \tilde{R}_{s-1}, O_{s-1}), (\rho_s, \tilde{R}_s \setminus \{b\}, \tilde{O}_s), \rho_s \cdot (W_s, b)\big).$$

*Case 2:* $\rho_s$ ends with a position of the cop player or $|\tilde{R}_s| = 0$.
We define
$$\zeta' = \big((\rho_1, \tilde{R}_1, O_1), \ldots, (\rho_{s-1}, \tilde{R}_{s-1}, O_{s-1}), \rho_s\big).$$

Now we prove that after the move of the robbers, all invariants that we have formulated still hold.

**Lemma 21.** *All invariants are preserved by the robbers' move.*

*Proof.* (Robs) holds by definition of the sets $\tilde{R}_i = R'_i$ and the construction of the memory update. (Lin) and (Cops) are obvious.

To prove (Omit), first notice that by (Omit) for $\zeta$, each set $O_i$ for $i = 1, \ldots, s-1$ is closed under reachability in $G - W_i$ and as, for $i = 1, \ldots, s-1$, we have $O'_i = O_i$ and $\rho'_i = \rho_i$, the invariant holds for all $i = 1, \ldots, s-1 \geq s'-2$. Moreover, $R'_i = \tilde{R}_i \subseteq O_i = O'_i$ holds by definition of the sets $\tilde{R}_i$ for $i = 1, \ldots, s-1$. In particular, in Case 2, there is nothing to show. So consider Case 1. There we have $s' = s+1$ and $O'_{s'-1} = O'_s = \tilde{O}_s = \text{Reach}_{G-W_s}(\tilde{R}_s \setminus \{b\})$ so $O'_s$ is obviously closed under reachability in $G - W_s$ and as $W'_{s'-1} = W_s$, $O'_s$ is closed under reachability in $G - W'_{s'-1}$. It remains to show that $R'_{s'-1} \subseteq O'_{s'-1}$. First we have $W_s \cap (\tilde{R}_s \setminus \{b\}) = \emptyset$. Indeed, assume that there is some $v \in W_s \cap (\tilde{R}_s \setminus \{b\})$. Then $v \notin U$ (as a cop and a robber cannot be on the same vertex) and according to (Cops) we have $U = \bigcup_{i=1}^{s} U_i$. So $v \notin U_s$ and hence, according to the definition of $U_s$, $v \in O^{s-1}$ which contradicts $v \in \tilde{R}_s$. So $R'_{s'-1} = \tilde{R}_s \setminus \{b\} \subseteq \tilde{O}_s = O'_{s'-1}$ and thus, (Omit) follows.

To prove (Ext), first notice that by (Ext) for $\zeta$, $O_i \subseteq \text{Reach}_{G-W_i^{-1}}(b_i)$ for $i = 1, \ldots, s-1$ and as $O'_i = O_i$ and $\rho'_i = \rho_i$ for $i = 1, \ldots, s-1$, the invariant holds for all $i = 1, \ldots, s-1 \geq s'-2$. In particular, in Case 2, there is nothing to show. Hence, we consider Case 1. First notice that $(W'_{s'-1})^{-1} = W_s^{-1} = \overline{W}_s$ and $b'_{s'-1} = b_s = \overline{b}_s$, so according to Lemma 19 we have $R'_{s'-1} \subseteq \tilde{R}_s \subseteq \text{Reach}_{G-W_s^{-1}}(b_s)$. Moreover, by definition, $O'_{s'-1} = \tilde{O}_s = \text{Reach}_{G-W_s}(\tilde{R}_s \setminus \{b\})$. So, if $v \in \tilde{O}_s$ then $v$ is reachable from some $\hat{b} \in \tilde{R}_s \setminus \{b\}$ in $G - W_s$ and as $\tilde{R}_s \subseteq \text{Reach}_{G-W_s^{-1}}(b_s)$, $\hat{b}$ is reachable from $b_s$ in $G - W_s^{-1}$. Thus, $v$ is reachable from $b_s$ in $G - (W_s^{-1} \cap W_s)$ and as $\rho_s = \hat{\rho}(W_s^{-1}, W_s, b_s)$ is consistent with $f$ by (Cons) for $\zeta$ and $f$ is monotone we have $v \in \text{Reach}_{G-W_s^{-1}}(b_s)$.

Finally, for (Cons), Case 2 is trivial. For Case 1, as $\rho_s$ is consistent with $f$ by (Cons), it suffices to show that $b \in \text{Reach}_{G-W_s^{-1}}(b_s)$. However, in Lemma 19 we have shown that $\tilde{R}_s \subseteq \text{Reach}_{G-\overline{W}_s}(\overline{b}_s)$ and as in Case 1 we have $\overline{b}_s = b_s$ and $\overline{W}_s = W_s^{-1}$, this follows from $b \in \tilde{R}_s$. □

Now we show that $\otimes_r f$ uses in fact at most $r \cdot k$ cops and by playing according to $\otimes_r f$, the cop player finally captures the robber. As we have already seen that $\otimes_r f$ is a monotone strategy, this concludes the proof of Theorem 10.

For the proof that $\otimes_r f$ uses at most $k \cdot r$ cops, first notice that by (Cops), the number of cops is bounded by $|\bigcup_{i=1}^{s} U_i|$. By definition of $U_i$ we have $|\bigcup_{i=1}^{s} U_i| \leq |\bigcup_{i=1}^{s} W_i|$. Due to (Cons), all $W_i$ have size at most $k$. It remains to show that there are at most $r$ distinct sets $W_i$.

**Lemma 22.** *For any memory state $\zeta$ which is consistent with $\otimes_r f$ we have $|\zeta| \leq r + 1$ and, if $|\zeta| = r + 1$ then $W_s = W_{s-1}$.*

*Proof.* In the following, we denote by $\zeta$ the memory state before the move of the cops, by $\zeta'$ the memory state after the move of the cops and before the move of the robbers and by $\zeta''$ we denote the memory state after the move of the robbers.

If $|\zeta| \leq r$ then obviously, $|\zeta''| \leq r + 1$. Consider the case where $|\zeta| = r + 1$ and $W_s = W_{s-1}$. As $|R| \leq r$, from (Robs) it follows that $R_i = \emptyset$ for some $i \in \{1, \ldots, s-1\}$ or $b_s \notin R$.

If $b_s \notin R$ then after the move of the cops we either have $|\zeta'| = r$ (if $R_{s-1} = \emptyset$) or we have $|\zeta'| = r + 1$ and $W'_{s'} = W'_s = W_{s-1} = W'_{s-1} = W'_{s'-1}$ (if $R_{s-1} \neq \emptyset$). Moreover, in that case the memory state after the move of the robbers (which is empty) is the same as after the move of the cops.

Now assume that $b_s \in R$ and let $i \in \{1, \ldots, s-1\}$ be such that $R_i = \emptyset$. Then, in the cops' move we are in Case II.1. If we are in Case II.1 (a) or in Case II.1 (c) then after the move of the cops we have $|\zeta'| = r$, so after the move of the robbers we clearly have $|\zeta''| \leq r + 1$. If we are in Case II.1 (b) then after the cops' move we have $s' = s$ and $\rho'_{s-1} = \rho_{s-1}$ and $\rho'_s = \rho_s$. Hence, $W'_{s-1} = W'_s$ and, as $\rho_s$ ends with a cops' position (because after the robbers' move $\rho_s$ always ends in a cops' position and Case II. (b) does not change $\rho_s$), $\zeta''$ is constructed according to Case 3 of the memory update after the move of the robbers. Hence, $|\zeta''| = |\zeta'| = r + 1$ and $W''_{s''} = W'_{s'} = W'_{s-1} = W''_{s''-1}$. □

To prove that $\otimes_r f$ is winning, we first prove the following additional property of this strategy. We use notation as above.

**(Progress)** For $i \in \{2, \ldots, s-1\}$, $R_i \cap O^{i-1} = \emptyset$ and $b_s \notin O^{s-1}$.

So (Progress) expresses that no history that we maintain induces to exclude vertices from placing cops which are currently occupied by robbers associated with greater histories. In particular, when when playing against the greatest robber, placing a cop on the vertices which is currently occupied by the robber will not be omitted. Notice that since $f$ is a winning strategy, at some point during any play which is consistent with $f$, the strategy $f$ will prescribe to place a cop on the vertex currently occupied by the robber. So, (Progress) tells us that we will not omit this cop-placement, which means, that (Progress) guarantees progress of the strategy $\otimes_r f$ against $r$ robbers.

Clearly, we also have to maintain this same property also for all smaller histories since, if at some point $b_s$ is either caught or returns to an earlier situation, one of the robbers associated with a history $\rho_i$ for some $i < s$ will become the greatest one.

We prove this invariant separately as (Progress) only uses the other invariants but is not further intertwined with them. Basically, (Progress) follows from the assumption that the robbers use an incomparably splitting strategy. However, as the sets $O_i$ are defined with respect to reachability in the graphs $G - W_i$, we have to transfer this topological incomparability from $G - U$ to the graphs $G - W_i$ for which we need that the robbers associated with the histories $\rho_i$ are compatible with these histories, see Lemma 11.

**Lemma 23.** *(Progress) is preserved by the move of the cops and the move of the robbers.*

*Proof.* First consider the situation after the move of the cops. In Case I we have $R'_j = R_j$ and $O_j = O'_j$ for $j = 1, \ldots, s-2$ and hence, $R'_j \cap (O^{j-1})' = \emptyset$ by

(Progress) for $\zeta$. Moreover, if $R_{s-1} = \emptyset$ then $s' = s - 1$, so $s' - 1 = s - 2$ and it remains to show that $b_{s'} \notin O^{s'-1}$. However, as $b_{s'} = b_s$ and $O^{s'-1} = O^{s-2}$ this follows immediately from (Progress) for $\zeta$.

If $R_{s-1} \neq \emptyset$ then $s' = s$ and $R'_{s-1} \subseteq R_s$ so $R'_{s-1} \cap (O^{s-2})' = \emptyset$ and $b'_s = b \notin (O^{s-2})'$ follows again immediately from $(O^{s-2})' = O^{s-2}$ and (Progress) for $\zeta$ and it remains to show that $b \notin O'_{s-1} = \tilde{O}_{s-1} = \text{Reach}_{G-W_{s-1}}(R_{s-1} \setminus \{b\})$. As the robbers use an isolating strategy, $b \notin \text{Reach}_{G-U}(R_{s-1} \setminus \{b\})$, now assume that $b \in \text{Reach}_{G-W_{s-1}}(R_{s-1} \setminus \{b\})$. Then, due to Lemma 12, $b \in \text{Reach}_{G-W^{s-1}}(R_{s-1} \setminus \{b\}) \subseteq \text{Reach}_{G-U^{s-1}}(R_{s-1} \setminus \{b\})$. Moreover, by Corollary 14, $\text{Reach}_{G-U^{s-1}}(R_{s-1} \setminus \{b\}) = \text{Reach}_{G-U}(R_{s-1} \setminus \{b\})$, which is a contradiction. In Case II, (Progress) for $\zeta'$ follows easily from (Progress) for $\zeta$ using the definition of the memory update.

Now consider the situation after the robbers' move. In Case 2, (Progress) holds by construction of the sets $\tilde{R}_i = R'_i$ for $i = 1, \ldots, s$. Moreover, in Case 1, $R'_i \cap (O^{i-1})' = \emptyset$ holds for $i = 1, \ldots, s' - 1$ by construction of the sets $R'_i$ as well and $b \notin (O^{s'-2})' = O^{s-1}$ holds by construction of $\tilde{R}_s$.

It remains to show that $b \notin O'_{s'-1} = \tilde{O}_s = \text{Reach}_{G-W_s}(\tilde{R}_s \setminus \{b\})$. As the robber player uses an isolating strategy we have $b \notin \text{Reach}_{G-U}(\tilde{R}_s \setminus \{b\})$, now assume that $b \in \text{Reach}_{G-W_s}(\tilde{R}_s \setminus \{b\})$. Then, as $W_s = W'_{s'-1}$ and $\tilde{R}_s \setminus \{b\} = R'_{s'-1}$, Lemma 12 for the memory state $\zeta'$ after the robbers' move yields $b \in \text{Reach}_{G-(W^{s'-1})'}(R'_{s'-1}) = \text{Reach}_{G-W^s}(\tilde{R}_s \setminus \{b\}) \subseteq \text{Reach}_{G-U^s}(\tilde{R}_s \setminus \{b\})$. Moreover, $U^s = U$, so $b \in \text{Reach}_{G-U}(\tilde{R}_s \setminus \{b\})$, which is a contradiction. $\square$

The following lemma concludes the proof of Theorem 10.

**Lemma 24.** $\otimes_r f$ *is winning.*

*Proof.* First observe that every cop that is placed on the graph according to the longest history restricts the reachability set of robber $b_s$, because these moves are according to $f$, which does not prescribe to make useless moves. Assume that there is some position in a play consistent with $\otimes_r f$ after which the reachability set of all robbers never becomes smaller (and at which the robbers are not caught yet). As it becomes smaller if one of the robbers moves (because the robbers play according to a prudent strategy), there must be point in time after which no robber moves. First we show, that if, in the move of the cops, Case II.2 happens infinitely often then $f$ is not winning, which contradicts our assumption.

So assume, that Case II.2 happens infinitely often. As the reachability set of $b_s$ never becomes smaller after some point in time, $\otimes_r f$ never places cops into $\text{Reach}_{G-U}(b_s)$ again. But $\text{Reach}_{G-W_s}(b_s) = \text{Reach}_{G-W^s}(b_s) \subseteq \text{Reach}_{G-U^s}(b_s) = \text{Reach}_{G-U}(b_s)$, so $\otimes_r f$ never places cops into $\text{Reach}_{G-W_s}(b_s)$ again. Since in this case $\otimes_r f$ places cops according to $f$ and every move that $f$ prescribes places cops into $\text{Reach}_{G-W_s}(b_s)$, $f$ prescribes to place cops only on vertices in $O^{s-1}$. Therefore, as this happens infinitely often, $b_s$ is never occupied by any cop according to $f$ due to the invariant (Progress). Hence, $f$ is not winning.

Now we show that Case I and Case II.1 can happen only finitely often, thus Case II.2 happens infinitely often. First, in Case I, either the number $s$ of histories in $\zeta$ or $|R_{s-1}|$ decreases. Moreover, in Case II.1, histories that are shorter than $\rho_s$ are extended or deleted (which decreases $s$), if they reach the next history. The length of the longest history in $\zeta$ is an upper bound for the growth of their lengths. Now as the robbers don't move, neither $s$ nor $|R_{s-1}|$ will ever increase again. Together, cases I and II.1 can happen only finitely many times. □

## 5 Robbers hierarchy and directed path-width

In this section we examine the hierarchy of complexity values for a given graph, induced by our new concept of graph searching:

$$\mathrm{dw}(G) = \mathrm{dw}_1(G) \leq \mathrm{dw}_2(G) \leq \ldots \leq \mathrm{dw}_n(G) = \mathrm{dpw}(G)$$

where $n$ is the number of vertices of $G$. We have already proved that for any $1 \leq r \leq n$, $\mathrm{dw}_r(G) \leq r \cdot \mathrm{dw}(G)$. Here we show that in general this hierarchy does not collapse which also proves that there is no bound on $\mathrm{dw}_r(G)$ just in terms of $\mathrm{dw}(G)$, independent of $r$. So in a sense, DAG-width can be approximated by a refinement of directed path-width and vice versa, but there are infinitely stages of approximation between those two measures.

**Theorem 25.** *For every $k > 0$, there is a class $\mathcal{G}^k$ of graphs such that for all $G \in \mathcal{G}^k$, $\mathrm{dw}_1(G) = 2 \cdot k$ and for all $r > 0$, there exists $G_r^k \in \mathcal{G}^k$ with*

1. $\mathrm{dpw}(G_r^k) = k \cdot (r+1)$, *where* dpw *denotes the directed path-width, and*
2. *for all $i \in \{1, \ldots, r\}$, $\mathrm{dw}_i(G_r^k) \geq \frac{i \cdot (k-1)}{2}$.*

*Proof.* Let $\oplus$ denote the lexicographic product of two graphs: for graphs $G = (V_1, E_1)$ and $H = (V_2, E_2)$, the lexicographic product $G \oplus H$ of $G$ and $H$ is a graph $(V_1 \times V_2, E')$ where $E'$ consists of pairs $((v_1, w_1), (v_2, w_2))$ with either $(v_1, v_2) \in E_1$, or $v_1 = v_2$ and $(w_1, w_2) \in E_2$.

The class $\mathcal{G}^k$ consists of graphs $G_r^k$, for each $r > 0$. Every $G_r^k = T_r \oplus K_k$ is the lexicographic product of the full undirected (i.e., with symmetrical edge relation) tree $T_r$ with branching degree $\lceil \frac{r}{2} \rceil + 2$ and of height[1] $r+1$, with the $k$-clique $K_k$. It is clear that $\mathrm{dw}_1(G_r^k)$ is $2 \cdot k$ (the cops play as on $T_r$ occupying the whole $K_k$-component instead of single tree vertices[2]). We have to show that $\mathrm{dpw}(G_r^k) = k(r+1)$ and that $\mathrm{dw}_i(G_r^k) \geq \frac{i \cdot (k-1)}{2}$.

We start with the path-width. A similar proof can be found, for example, in [4]. The common intuition is that the cops clean the graph contaminated by the robber. Note that the branching degree of all $T_r$ is at least 3. We prove that $\mathrm{dpw}(T_r) = r + 1$, the statement with factor $k$ follows as for DAG-width. The proof is done by induction on $r$. The case $r = 1$ is trivial. If $r + 1$ cops win on $T_r$ then $r + 2$ cops win on $T_{r+1}$ by placing a cop on the root and applying the strategy for $r + 1$ cops from the induction hypothesis for every subtree.

The other direction (that $\mathrm{dpw}(T_r) \geq r + 1$) is also proven by induction on $r$. The induction base is clear. Assume that $\mathrm{dpw}(T_r) \geq r+1$. In $T_{r+1}$, let the direct successors of the root be $v_1, \ldots, v_m$ (notice that $m \geq 3$). All subtrees $T^i$ rooted at $v_i$, for $i \in \{1, \ldots, m\}$ must be decontaminated and $r+1$ cops are needed for that. Assume w.l.o.g. that $T^1$ is decontaminated first. Later on, the other subtrees must be decontaminated. Assume that the first of them is $T^2$. Again, all $r+1$ cops are needed for that, i.e., in some position they all are in $T^2$. But then there is a path from $T^m$ via the root of the whole tree to $T^1$. Thus $T^1$ becomes recontaminated, which contradicts the monotonicity of path-width.

---
[1] We define the height of the tree such that a single vertex has height 1.
[2] The idea to use the lexicographic product and of the proof is due to [7].

It remains to show that $k \cdot i$ robbers win against $\frac{i \cdot (k-1)}{2}$ cops on $G_r^k$. We show only that $i$ robbers win against $\lfloor \frac{i}{2} \rfloor$ cops on $T_r$, the result with factor $k$ follows as above. As in the proof of Theorem 7 we can assume the robbers play top-down.

The winning strategy of robbers is to bound every cop. A cop is *bounded* if there is a cop free path from a robber to the cop. When a cop is placed on a vertex $v$, the robbers occupy two subtrees of $v$. As there are at least two robbers for each cop, this is always possible. At the latest when a cop reaches level $\lfloor \frac{i \cdot (k-1)}{2} \rfloor$ (counting from the root), all cops are bounded. □

## Acknowledgements

We thank Łukasz Kaiser for many inspiring discussions.